%% file: main.tex
\documentclass[camera,10pt]{jpaper} 
\DeclareMathAlphabet{\mathcal}{OMS}{cmsy}{m}{n}
\usepackage{setspace} 
\usepackage[italic]{mathastext}
\usepackage{array}
\usepackage{balance}
\newcommand{\ignore}[1]{}
\usepackage{fancyhdr}
\usepackage[normalem]{ulem}
\usepackage{datetime} 
\usepackage[binary-units=true]{siunitx}
\usepackage{booktabs}
\usepackage{dblfloatfix}
\usepackage{hyperref}
\usepackage{multirow}
\usepackage{multicol}
\usepackage{float}
\usepackage[shortlabels]{enumitem}
\usepackage{xurl}
\usepackage{cite}
\usepackage[resetlabels,labeled]{multibib}
\usepackage{xcolor}
\usepackage[font={small,bf}]{caption}
\usepackage{pifont}
\usepackage{amssymb}
\usepackage{amsmath}

\definecolor{dollarbill}{rgb}{0.52, 0.73, 0.4}

\newcolumntype{L}[1]{>{\raggedright\let\newline\\\arraybackslash\hspace{0pt}}m{#1}}
\newcolumntype{C}[1]{>{\centering\let\newline\\\arraybackslash\hspace{0pt}}m{#1}}
\newcolumntype{R}[1]{>{\raggedleft\let\newline\\\arraybackslash\hspace{0pt}}m{#1}}

\setlist[itemize]{noitemsep,itemsep=0pt,parsep=0pt,topsep=0pt,partopsep=0pt,leftmargin=1em}
\setlist[enumerate]{noitemsep,itemsep=0pt,parsep=0pt,topsep=0pt,partopsep=0pt,leftmargin=1em}



\definecolor{grey}{rgb}{0.7, 0.6, 0.6}
\definecolor{amber}{rgb}{1.0, 0.49, 0.0}
\definecolor{darkamber}{rgb}{0.9, 0.49, 0.0}
\definecolor{extradarkamber}{rgb}{0.55, 0.295, 0.0}
\definecolor{darkgreen}{rgb}{0.0, 0.2, 0.13}
\definecolor{darkbyzantium}{rgb}{0.36, 0.22, 0.33}
\definecolor{darkseagreen}{rgb}{0.56, 0.74, 0.56}
\definecolor{darkspringgreen}{rgb}{0.09, 0.45, 0.27}
\definecolor{dollarbill}{rgb}{0.52, 0.73, 0.4}
\definecolor{darkcerulean}{rgb}{0.03, 0.27, 0.49}

\newif\ifcameraready
\camerareadytrue

\ifcameraready

\else

\fi

\ifcameraready
  \usepackage[disable]{todonotes}
   
  \else
  \usepackage[textsize=scriptsize]{todonotes}
   
\fi

\makeatletter
\g@addto@macro{\normalsize}{%
  \setlength{\abovedisplayskip}{2pt plus 1pt minus 1pt}
  \setlength{\belowdisplayskip}{2pt plus 1pt minus 1pt}
  \setlength{\abovedisplayshortskip}{0pt}
  \setlength{\belowdisplayshortskip}{0pt}
  \setlength{\intextsep}{4pt plus 1pt minus 1pt}
  \setlength{\textfloatsep}{4pt plus 1pt minus 1pt}
  \setlength{\skip\footins}{5pt plus 1pt minus 1pt}}
\makeatother

\titlespacing\section{0pt}{2pt plus 1pt minus 1pt}{3pt plus 1pt minus 2pt}
\titlespacing\subsection{0pt}{2pt plus 1pt minus 1pt}{3pt plus 1pt minus 2pt}
\titlespacing\subsubsection{0pt}{2pt plus 1pt minus 1pt}{3pt plus 1pt minus 2pt}

\usepackage{tikz}
\usepackage{fancyhdr}
\usepackage{datetime} 

\settimeformat{ampmtime}

\newif\ifuseversions
\useversionsfalse
\newcommand{\versionnum}[0]{26}
\ifuseversions
  \def\parsepdfdatetime#1:#2#3#4#5#6#7#8#9{%
    \def\theyear{#2#3#4#5}%
    \def\themonth{#6#7}%
    \def\theday{#8#9}%
    \parsepdftime
  }

  \def\parsepdftime#1#2#3#4#5#6#7\endparsepdfdatetime{%
    \def\thehour{#1#2}%
    \def\theminute{#3#4}%
    \def\thesecond{#5#6}%
    \ifstrequal{#7}{Z}
    {%
      \def\thetimezonehour{+00}%
      \def\thetimezoneminute{00}%
    }%
    {%
      \parsepdftimezone#7%
    }%
  }

  \def\parsepdftimezone#1'#2'{%
    \def\thetimezonehour{#1}%
    \def\thetimezoneminute{#2}%
  }

  \newcommand*{\thetimezone}{\thetimezonehour:\thetimezoneminute}
  \expandafter\parsepdfdatetime\pdfcreationdate\endparsepdfdatetime

  \settimeformat{ampmtime}
  \newcommand{\version}[1]{\emph{Version #1 (Built:~\today~@ \currenttime~UTC\thetimezone)}}
  \AtBeginShipout
  {\AtBeginShipoutAddToBox{
      \begin{tikzpicture}[overlay, remember picture]
      \node[anchor=north] at (current page.north) {\textcolor{blue}{\vspace{3em}\version{\versionnum}}};    
      \end{tikzpicture}
  }}
\fi

  \ifcameraready
\else
  
  \fancypagestyle{firststyle}
  {
    \fancyfoot[C]{\thepage}
  }
\fi

\newcites{S}{Survey Sources}

\usepackage{xspace}
\newcommand{\pim}{PIM\xspace}
\newcommand{\pimlong}{Processing-In-Memory\xspace}
\newcommand{\cim}{CIM\xspace}
\newcommand{\cimlong}{Computation-In-Memory\xspace}
\newcommand{\mechanism}{Demeter\xspace}
\newcommand{\accmechanism}{Acc-Demeter\xspace}
\newcommand{\cmechanism}{C-Demeter\xspace}
\newcommand{\gmechanism}{G-Demeter\xspace}
\newcommand{\hdrefs}{HD-RefDB\xspace}
\newcommand{\hdreads}{HD-ReadDB\xspace}
\newcommand{\hdc}{HDC\xspace}
\newcommand{\hd}{HD\xspace}
\newcommand{\ngram}{N-gram\xspace}
\newcommand{\ngrams}{N-grams\xspace}

\newcommand{\krakennew}{Kraken2\xspace}

\newcommand{\kbr}{Kraken2+Bracken\xspace}

\newcommand{\metacache}{MetaCache\xspace}
\newcommand{\clark}{CLARK\xspace}
\newcommand{\am}{AM\xspace}
\newcommand{\amlong}{Associate Memory\xspace}
\newcommand{\im}{IM\xspace}
\newcommand{\imlong}{Item Memory\xspace}
\newcommand{\bbopinit}{\emph{bbop\_init}\xspace}
\newcommand{\bbopop}{\emph{bbop\_op}\xspace}
\newcommand{\src}{\emph{src}\xspace}
\newcommand{\dst}{\emph{dst}\xspace}

\newcommand\fig[1]{Fig.~{#1}\xspace}
\newcommand\figs[1]{Figures~{#1}\xspace}
\newcommand\sect[1]{Section~{#1}\xspace}
\newcommand\sects[1]{Sections~{#1}\xspace}

\newcommand\eque[1]{Equation~{#1}\xspace}

\newcommand\tab[1]{Table~{#1}\xspace}

\newcommand*\circled[1]{\tikz[baseline=(char.base)]{
            \node[shape=circle,draw,inner sep=0pt,fill=black, text=white] (char) {#1};}}
\newcommand*\Rcircled[1]{\tikz[baseline=(char.base)]{
            \node[shape=circle,draw,inner sep=0pt,fill=red, text=white] (char) {#1};}}

\begin{document}

\bstctlcite{IEEEexample:BSTcontrol}
\bstctlcite[@auxoutS]{IEEEexample:BSTcontrol}

\title{\mechanism : A Fast and Energy-Efficient Food Profiler \\using Hyperdimensional Computing in Memory}

\newcommand{\tud}{{\large$^\dagger$}} 
\newcommand{\ethz}{{\large$^\ddagger$}}
\newcommand{\scomma}{{\large$^,$}}

\author{ \vspace{-2ex}\\%
Taha Shahroodi\tud{}\hspace{0.15in}%
Mahdi Zahedi\tud{}\hspace{0.15in}%
Can Firtina\ethz{}\hspace{0.15in}%
Mohammed Alser\ethz{}\hspace{0.15in}\\%
Stephan Wong\tud{}\hspace{0.15in}%
Onur Mutlu\ethz{}\hspace{0.15in}%
Said Hamdioui\tud{}\vspace{2mm}\\%
\textit{\tud{}TU Delft\hspace{0.30in}\ethz{}ETH Z{\"u}rich}%
\\}

\maketitle

\newcommand\blfootnote[1]{%
  \begingroup
  \renewcommand\thefootnote{}\footnotetext{#1}%
  \addtocounter{footnote}{-1}%
  \endgroup
}

\begin{abstract}
  \input{Sections/0_abstract}

\end{abstract}  

\input{Sections/1_introduction}

\input{Sections/2_background}

\input{Sections/3_00_framework}

\input{Sections/4_EvaluationDemeter}

\input{Sections/5_00_accelerator}

\input{Sections/6_SystemIntegration_DataLayout_ISA_ProgrammingInterface_ComputationOnLargeData}

\input{Sections/9_00_EvaluationAccelerator}

\input{Sections/7_LimitationOrDiscussions}

\input{Sections/10_related}

\input{Sections/11_conclusion}

\section*{Acknowledgments}

We thank the members of the SAFARI Research Group at ETH Zurich and the Quantum \& Computing Engineering at TU Delft for their valuable feedback and the constructively critical environment that they provide. We specifically thank Abhairaj Singh for his feedback regarding PCM devices and circuit designs. We acknowledge the generous gifts provided by our industry partners, including Google, Huawei, Intel, Microsoft, and VMware. We acknowledge support from the ETH Future Computing Laboratory and the Semiconductor Research Corporation.

\setbiblabelwidth{1000} 
\bibliographystyle{IEEEtran}
\balance
\bibliography{main}



\end{document}

%% file: Sections/0_abstract.tex
Food profiling is an essential step in any food monitoring system needed to prevent health risks and potential frauds in the food industry. Significant improvements in sequencing technologies are pushing food profiling to become the main computational bottleneck. State-of-the-art profilers are unfortunately too costly for food profiling. Our goal is to design a food profiler that solves the main limitations of existing profilers, namely (1) working on massive data structures and (2) incurring considerable data movement for a real-time monitoring system. To this end, we propose \mechanism, the first platform-independent framework for food profiling. \mechanism overcomes the first limitation through the use of hyperdimensional computing (\hdc) and efficiently performs the accurate few-species classification required in food profiling. We overcome the second limitation by using an in-memory hardware accelerator for \mechanism (named \accmechanism) based on memristor devices.  \accmechanism actualizes several domain-specific optimizations and exploits the inherent characteristics of memristors to improve the overall performance and energy consumption of \accmechanism. We compare \mechanism's accuracy with other industrial food profilers using detailed software modeling. We synthesize \accmechanism's required hardware using UMC's 65nm library by considering an accurate PCM model based on silicon-based prototypes. Our evaluations demonstrate that \accmechanism achieves a (1) throughput improvement of 192$\times$ and 724$\times$ and (2) memory reduction of 36$\times$ and 33$\times$ compared to \krakennew and \metacache (2 state-of-the-art profilers), respectively, on typical food-related databases. Demeter maintains an acceptable profiling accuracy (within 2\% of existing tools) and incurs a very low area overhead.

%% file: Sections/1_introduction.tex
\section{Introduction}
\label{sec:introduction}

The urgent need for a real-time, efficient, and accurate food monitoring system is apparent when one considers the economic impacts and health risk issues due to human errors and/or intentional fraud regarding everyday food. For example, a worldwide annual loss of \$10 to \$24 billion is estimated only for the frauds happening in the fish industry~\cite{agnew2009estimating}. The Halal meat scandal~\cite{smith2004rural} and the black fish scandal~\cite{smith2015documenting} are just a few other preventable examples that could have been quickly prevented if we had accurately and efficiently monitored all the food productions in real-time.

Food profiling is the first step and the only computationally expensive task in a food monitoring system. The food profiling task entails determining the existing species in a food sample and their relative abundance~\cite{AFSMetacache, ripp2014all}. Today's food profilers work with sequenced data as we can capture a more accurate profile using the sequences of a food sample. The rapid drop in the cost of DNA sequencing in the past decades and the expectation for a continual trend~\cite{DNAdropCost2020, DNAdropCost2013}\footnote{Researchers expect that soon different analyses of the sequenced data will become the cost and performance bottleneck rather than sequencing itself, as is currently the case for read mapping vs. DNA sequencing~\cite{alser2020accelerating}.} is expected to lead the way for profiling to become the main bottleneck of this pipeline.

Currently, the industry utilizes state-of-the-art (SOTA) taxonomic profilers from metagenomic studies for food profiling due to the similarity of problem statements in food profiling and metagenomics profiling. However, as alluded to, such profilers are developed as the first step of metagenomic studies~\cite{handelsman1998molecular, rondon2000cloning, wooley2010primer}: a new yet different line of research that allows us to study many species that are taken directly from their environment altogether, as opposed to studying them individually. Unfortunately, these profilers are overkill for simply profiling a given food sample and, therefore, costly since those taxonomic profilers have been designed for different, more complex goals such as (1) capturing complex operations between organisms or (2) finding insights on species that cannot be clonally cultured in labs. Such profilers are also designed for working on larger, more complex, and randomly mixed genome sequences and demand a significant amount of resources that simply impede real-time monitoring of all food samples after production, shipment, or distribution; the ultimate goal of a food monitoring system. Therefore, a new solution must be sought for food profiling that is cheaper, faster, more energy-efficient, and yet accurate.

In particular, we pinpoint two critical sources of inefficiency in SOTA profilers currently used for food monitoring, collectively called food profilers or profilers hereafter. First, all current (food) profilers work with significantly large working data structures, e.g., humongous hash tables or sorted lists, that require high-end servers with extensive storage and memory capabilities to be handled. This fundamentally limits performance scaling on par with that in sequencing technologies. Second, current profiling techniques incur a significant number of random accesses to large working datasets, and as a result, unnecessary data movement between their storage and memory plus their memory and compute units which cannot be otherwise done where the data resides due to (1) the size of the final data structures and (2) the required operations for tasks in hand. This directly translates to massive energy consumption and latency. For example, as shown in our evaluations \sects{\ref{sec:evaluation_Demeter}, \ref{sec:evaluation}}, a widely used SOTA profiler takes $\sim$1 minute to profile one high-coverage sequenced food sample. However, it requires a super machine or cluster with at least \SI{300}{\giga\byte} of memory and proportionally scaled-up compute power. These costs add up to an unbearable required time and equipment for real-time monitoring of all existing and producing food samples. Therefore, a healthy economy regarding the food industry cannot keep using these profilers and demands cheaper, faster, more energy-efficient, and accurate food profilers for the years to come.

\textbf{Our goal} in this work is to solve both limitations of previous profilers, namely (1) reliance on high-end servers and scaling problems due to required massive data structures and (2) incurring unnecessary data movement. \textbf{To this end,} we propose \mechanism, an end-to-end, hardware/software co-designed food profiling framework that efficiently profiles species of a food sample. \textbf{The key idea} of \mechanism is to reduce the food profiling problem to a multi-object (multi-species) classification problem using hyperdimensional (\hd) computing (\hdc) followed by an abundance estimation step. \mechanism is a platform-independent framework and produces accurate results on any hardware platform such as a central processing unit) CPU, graphics processing unit (GPU), or application-specific integrated circuit (ASIC).

Our experiments show that although the accuracy of \mechanism is comparable with existing SOTA profilers, typical processing units (CPUs) are not exploiting the full parallelism offered by our \hdc-based approach, prohibiting those platforms from outperforming SOTA profilers. Moreover, we find two more optimization opportunities that can be achieved with a wisely-chosen platform: (1) eliminating the cost of existing shift operations and (2) mitigating the significant amount of data movement involved in our \hdc-based solution. Therefore, we propose an in-memory hardware accelerator for \mechanism, \accmechanism, to mitigate the costs mentioned above and simultaneously solve the second problem of profilers as well. \accmechanism achieves these by (1) the physical attributes of nanoscale memristive-based devices\footnote{We choose Phase Change Memory (PCM) devices as members of memristor families due to our accessibility to accurate measurements and models. However, in principle, our proposed techniques can be applied to any memristor-based memory technology, such as ReRAM or STT-MRAM.}, (2) \pimlong (\pim), where the data resides, and (3) zero-overhead shift operation in hardware. It is worth noting that, with the advent of portable sequencing machines, a move from cloud computing with sophisticated infrastructure towards an in-build profiler (or other genomics-related kernels) inside the sequencer is finally in the foreseeable future.

Our paper makes the following main contributions:
\begin{itemize} 
    \item To our knowledge, \mechanism is the first framework that enables food profiling via \hdc. \mechanism provides a five-step approach to determine the relative abundance of a set of the food read sequences at the species level. We design \mechanism to (1) address the key problems of food profiling rather than accelerating regular metagenomic profilers and (2) be platform-independent (\sect{\ref{sec:framework}}). 

    \item We propose a \pim-enabled hardware accelerator for \mechanism using memristor devices (\accmechanism) to extract \mechanism's full potentials and solve the data movement problem in \mechanism and previous profilers. We propose several optimization techniques for \accmechanism based on domain-specific knowledge of food profiling and our background in PCM cells characteristics and \hdc operations. To our knowledge, \accmechanism is the first (in-memory) hardware accelerator for a food profiler (\sect{\ref{sec:accelerator}}).  

    \item We rigorously compared \mechanism and \accmechanism to four SOTA food profilers. We show that \mechanism provides an accuracy level comparable with previous food profilers and within the accepted level of food monitoring systems. The default setting of \accmechanism enables a (1) throughput improvement of $\sim$192$\times$ and 724$\times$ and (2) reduction in the required memory of $\sim$36$\times$ and 33$\times$ compared to \krakennew~\cite{kraken2} and \metacache~\cite{AFSMetacache}, respectively, when querying on a typical food-related reference genome database, i.e., AFS20~\cite{AFSMetacache}. Our design requires only $\sim$8.9 $mm^2$ die area and can process $\sim$9.45 Mbp per joule for our largest food-related database AFS31~\cite{AFSMetacache} (\sect{\ref{sec:evaluation}}). 
    
\end{itemize}

%% file: Sections/2_background.tex
\section{Background and Motivation} \label{sec:background}

This section discusses the necessary background and introduction to (1) the current taxonomic profilers and their shortcomings when used for food profiling, (2) \hdc, and (3) the \pim paradigm. We devote the materials mainly to those closely related to or used by \mechanism. For more detailed background information, we refer the reader to comprehensive reviews on these topics~\cite {perez2020metagenomic, mcintyre2017comprehensive, breitwieser2019review, ge2020classification, kleyko2018classification, nguyen2020classification, hamdioui2015memristor}.

\subsection{Metagenomic Profilers}
\label{subsec:profilers_background}

Constantly increasing the performance of sequencing technologies and the fast drop in the cost of DNA sequencing~\cite{DNAdropCost2013, DNAdropCost2020} catalyzed the metagenomic studies~\cite{handelsman1998molecular, rondon2000cloning, wooley2010primer}. These studies enable us to capture the big picture of the environment without isolating or cultivating individual organisms. For this purpose, one needs to perform taxonomic profiling: determining the relative abundances of species in a sample directly taken from the environment. Due to the high cost associated with alignment and assembly for large reference datasets, to this date, we still prefer heuristics statistical-based profilers to assembly- or alignment-based ones. However, even these profilers are not yet cheap or economic and prevent large-scale, real-time studying. Their cost is mainly related to the required memory for profilers' data structure and algorithms. Such large data structures or sophisticated algorithms force us to use high-end servers and are needed to fulfill complex goals of subsequent metagenomic analysis, namely capturing complex operations between organisms and discovering insights on species that can not be clonally cultured in labs. This high cost of profiling in metagenomics profiler prevents us from efficiently profiling food samples in real-time, the end goal of a food monitoring system.

\subsection{Problems of Food Profilers}
\label{subsec:problems_food_profiler_largeDataStructure_L2L3misses_background}

We use VTune~\cite{IntelVTune} and profile three SOTA profilers that are currently used for food samples as well, namely \krakennew, \clark, and \metacache, using their default datasets and parameters on the original platforms for which they have been designed. We make two main observations, which follow a similar trend reported in previous studies in genomics as well~\cite{wu2021sieve, kraken2, zhou2021ultra}.

\noindent
\textbf{Observation 1.} All these profilers induce large memory requirements for their data structures. For example, \krakennew requires a minimum of \SI{300}{\giga\byte} memory for its reference data structure. Even for smaller and less complex reference data bases such as those in food industry, \krakennew still requires more than \SI{50}{\giga\byte} of memory (\sect{\ref{subsec:storageandMemory_evaluation}}).

\noindent
\textbf{Observation 2.} All profilers induce high miss rates in L2 and L3 ($\sim$68 to 90\%). The nature of their underlying algorithms causes this inefficiency because they always query a small fraction of keys in a large hash table and/or sorted list, leading to random memory access patterns. In other words, the arithmetic intensity of the profilers is too small to the extent that even increasing the number of threads does not help resolve the CPU stall cycles caused by memory accesses required by these misses.

Overall, current food profilers' large working data structure and their low arithmetic intensity lead to high storage cost, low performance, and high energy consumption. It also demands high-end servers. This motivates designs (such as \mechanism and \accmechanism) that provide reduced working data structures, eliminate unnecessary data movements, and can liberate us from dependency on the clusters.

\subsection{Hyperdimensional Computing}
\label{subsec:HDC_background}

Hyperdimensional computing (\hdc)~\cite{kanerva2009hyperdimensional, gayler2004vector} is a brain-inspired computing paradigm that has been demonstrated to be effective in reference-based learning domains, such as text classification~\cite{najafabadi2016hyperdimensional, kleyko2015fault, kleyko2015imitation}, gesture recognition~\cite{rahimi2016hyperdimensional}, and latent semantic analysis~\cite{kanerva2000random}. Elements of \hdc are presented using high-dimensional vectors, hereafter called \hd vectors. \hd vectors can be composed of real~\cite{plate1995holographic, gayler1998multiplicative, gallant2016positional}, binary~\cite{kanerva2009hyperdimensional, rachkovskij2001representation}, bipolar~\cite{gallant2013representing, rahimi2016hyperdimensional}, or complex numbers~\cite{plate2003holographic}. Previous works show that binary representations of \hd vectors are more practical and efficient for classification problems or one-shot reinforcement learning. This representation is also more hardware-friendly. Therefore, we proceed with binary \hd representations. \hd vectors also come with other powerful features such as robustness to random errors, holistic representation, and randomness. We refer the enthusiastic readers to previous works for more details on these features~\cite {kanerva2009hyperdimensional, kleyko2018classification}.

Like other reference-based classifiers, an \hdc-based system also takes two steps: (1) training and (2) classification. An encoding mechanism is used in both steps. One famous example is the \ngram encoding mechanism that follows a two-step approach for encoding a string of size L to an \hd vector of size D. \textbf{Step~1:} It combines N consecutive alphabets of the string and builds an \hd vector that is orthogonal to them all and can preserve their relative order. This operation is called binding and is represented in \eque{\ref{eq:ngram_xorshift}}, where $\rho^i (X)$ represents the $i^{th}$ permutation of vector X and $B_i$ are once randomly-generated representative \hd vectors (also referred to as atomic or basis \hd vectors) for the $i^{th}$ character of the string $C_i$. The string is a DNA sequence in \mechanism.

\begin{equation}
\ngram (C_1, C_2, ... , C_N) = B_1 \oplus \rho (B_2) \oplus \dots \oplus \rho ^{N-1}(B_N)
\label{eq:ngram_xorshift}
\end{equation}

\textbf{Step~2:} The encoder performs an element-wise addition between all \hd vectors corresponding to consecutive \ngrams, called bundling, to present the entire input sequence. To binarize the final \hd vector, the encoder applies a majority function over each position. This final vector is stored in associate memory (AM) and is called a prototype \hd vector if the input was a reference genome. Otherwise, it is called query \hd vector and will use it for classification.

The most common approach for classifying whether the sequence query belongs to any of the classes in AM after using \ngram encoding mechanism is to measure the hamming distance between the query \hd vector (Q) and each of the prototype \hd vectors (Ps) and decide based on a fixed distance or threshold (T). This can be easily performed with an XNOR of Q and each P followed by a pop-count\footnote{Pop-count (population count) of a vector or specific value is the process of finding the number of set bits (1s) in that value.} and thresholding operation, as shown in \eque{\ref{eq:classification}}.

\begin{equation}
 Classification(i)= 
   \begin{cases}
     1,\,\,\,\,\, \text{if}\sum\limits_{j=1}^{D}Q(j)\bar\oplus P_{i}(j) \geq T\\
     0,\,\,\,\,\, \text{otherwise}
   \end{cases}
   \label{eq:classification}
\end{equation}

\subsection{Computing Inside/Near Memory} \label{subsec:pimcim_background}

For decades, the processing units have been developed at a faster rate than memory units, causing memory units to become a bottleneck, especially in data-intensive workloads. The \pimlong (\pim) (interchangeably also referred to as \cimlong (\cim)) is a promising paradigm that aims to alleviate the data movement bottleneck. In essence, \pim advocates for avoiding unnecessary data movement and redesigning systems such that they are no longer processor-centric. Previous works show the potential of various memory technologies for implementing \pim-based architectures~\cite{seshadri2017ambit, lee2010phase, kang2018multi, chen2010advances}. Resistive memories or memristive devices, such as ReRAM and PCM~\cite{waser2009redox, lee2010phase, HDC-CIM-IBM}, have recently been introduced as a suitable candidate for both storage and computation units that can efficiently perform vector-matrix multiplication~\cite{xia2019memristive} and bulk bit-wise logical operations~\cite{xie2017scouting, cheng2019functional} since they can follow Kirchhoff’s law inherently. They also enjoy non-volatility, high-density, and near-zero standby power. Due to the inherent high parallelism, simplicity of the required operations, and intrinsic robustness and error tolerance of an \hdc-based system, such a system fits well with the \pim paradigm. A few recent works propose application-specific hardware accelerators based on memristive devices~\cite{ielmini2018memory, li2016hyperdimensional, wu2018brain, HDC-CIM-IBM} for such \hdc-based systems. We describe the differences between our implementation and closest previous proposals in \sects{\ref{sec:accelerator} and~\ref{sec:relatedwork}}.

%% file: Sections/3_00_framework.tex
\section{\mechanism} \label{sec:framework}

\begin{figure*}[t]
\centering
    \includegraphics[width=1\linewidth]{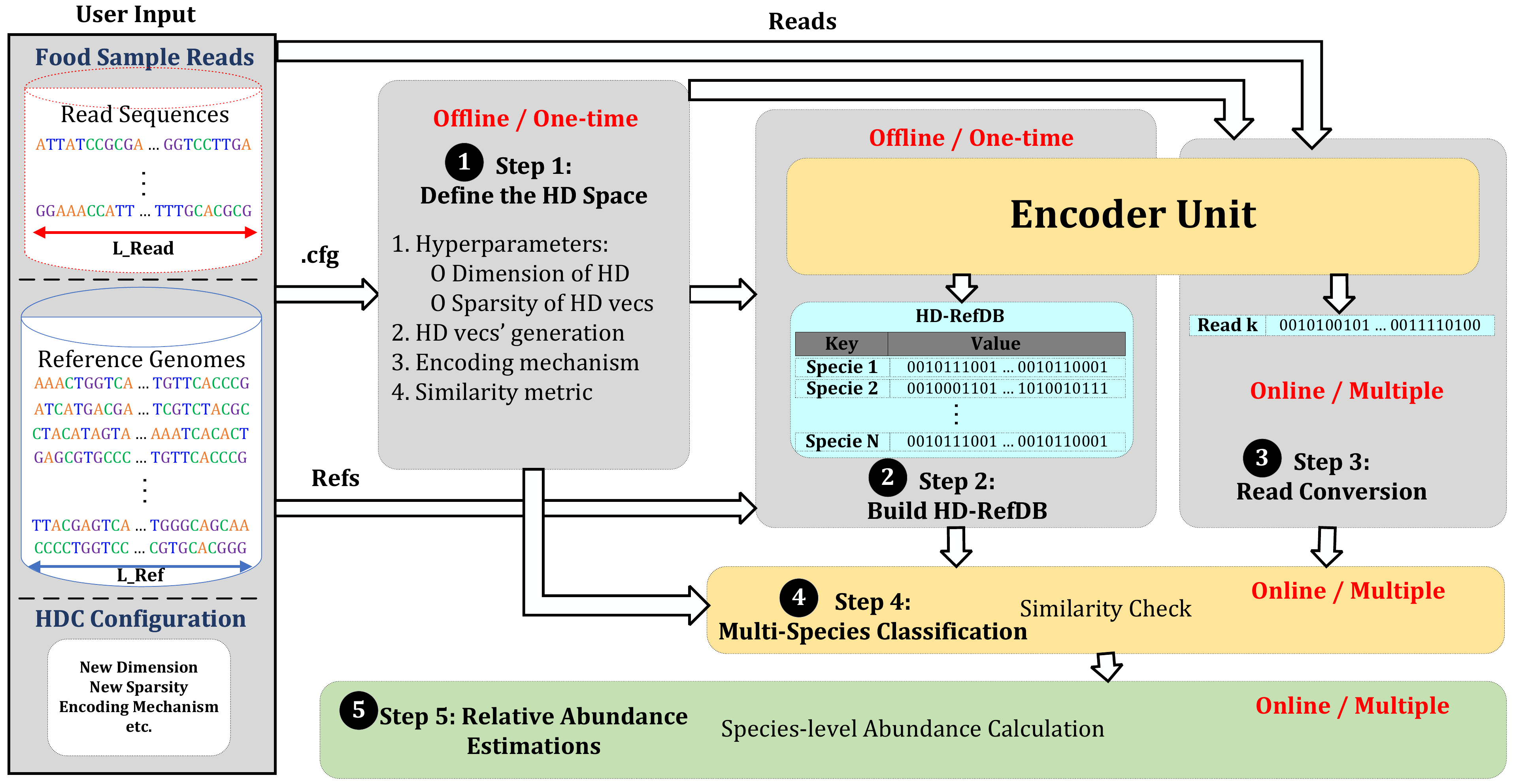}
    \caption{Overview of \mechanism framework.}
    \label{fig:overview_framework}
\end{figure*}

\mechanism is a configurable framework for food profiling and is based on three main insights: (1) current food profilers, whereas accurate, are neither memory- nor energy-efficient, (2) the primary sources of high cost and inefficiency in current food profilers is their large reference data structures and working sets, and (3) one can profile food samples quickly and accurately using \hdc. \fig{\ref{fig:overview_framework}} provides an overview of the five key steps in \mechanism: \circled{1} defining an \hd space, \circled{2} building an \hd reference database (\hdrefs), \circled{3} converting sample reads into \hd space, \circled{4} determining the possible species assignment per sample read, and \circled{5} performing abundance estimation. We describe each step in more detail next.

\input{Sections/3_1_HDSpace}

\input{Sections/3_2_DBHDRef}

\input{Sections/3_3_HDRead}

\input{Sections/3_4_similaritycheck}

\input{Sections/3_5_relativeabundance}

%% file: Sections/3_1_HDSpace.tex
\subsection{Step~1: Define the HD Space} \label{subsec:hdspace_framework}

As the first step (\circled{1} in \fig{\ref{fig:overview_framework}}), \mechanism defines an \hd space for all subsequent operations and steps. This is a crucial step as it determines the operations in the remaining steps. Unfortunately, many previous \hdc-based proposals did not support the user's input for determining the \hd space and designed their space statically. Hence, such designs are more limited.

\mechanism defines the \hd space in 4 stages. \textbf{Stage~1:} \mechanism fixes two hyperparameters: (1) The dimension of the \hd space; i.e., the dimensionality of the \hd vectors (element representations), and (2) The sparsity of each element (\hd vector). \textbf{Stage~2:} \mechanism generates a few atomic \hd vectors and stores them in memory (commonly called \imlong (\im)). These vectors can be (1) the \hd vectors that represent our genome alphabets or (2) the one-time randomly generated \hd vectors that some encoding mechanisms use, for example, to introduce the concept of order between alphabets of one input. \textbf{Stage~3:} \mechanism decides on the encoding mechanism to build the space with. This very encoding mechanism will be used throughout Steps~\circled{2} and \circled{3} of \mechanism. \textbf{Stage 4:} \mechanism fixes the similarity metric and any other associated parameters (such as thresholds) based on the user's input or a common choice considering previous stages. \mechanism stores a default value for each stage in a configuration file. Once the user summons \mechanism, \mechanism quickly checks if a configuration file matches with the user's requested \hd space or not. \mechanism only runs this step if such file does not exist or the user asked for a change.

%% file: Sections/3_2_DBHDRef.tex
\subsection{Step~2: Build \mechanism's Reference Data Structure} \label{subsec:DBHDRef_framework}

\mechanism takes two sets of inputs in step~\circled{2}: (1) \hd space parameters defined in Step~\circled{1} and (2) a reference genome database. Subsequently, \mechanism builds a new reference database in its \hd space out of all the considered reference genomes. This new database, called \hdrefs, consists of one (or few) prototype \hd vector(s) from any given reference genome in the original reference database and is stored in AM. \hdrefs can be as varied as the number of combinations of possible hyperparameters, atomic vectors, and encoding mechanisms in Step~\circled{1}. This step aims to reduce the size of the working set for the classification task while avoiding accuracy drop. Since this step requires only simple arithmetics and is also highly parallelizable, it can still be accelerated on our proposed \pim-enabled accelerator (\sect{\ref{sec:accelerator}}).

%% file: Sections/3_3_HDRead.tex
\subsection{Step~3: \mechanism's Read Conversion} \label{subsec:HDRead_framework}

\mechanism again takes two inputs in Step~\circled{3}: (1) \hd space configuration and (2) read sequences of the food sample under study. \mechanism translates each of these read sequences into one query \hd vector. To prevent any extra storage cost and to pipeline computations of Step~\circled{3} and Step~\circled{4}, \mechanism forwards each query \hd vector to the next step instead of storing them inside a memory unit while waiting for all of them to be constructed first\footnote{This is the default behavior in \mechanism. However, we also provide the option for the user to keep and store these query \hd vectors (\hdreads) in case one needs to analyze them further. If one uses this option, the stored query \hd vectors create another database representing the reads in our food sample, called \hdreads hereafter.}. Query \hd vectors created in this step can require larger or smaller space than a read, depending on the initial length of the read sequences and the dimension of the \hd space. Therefore, although Steps~\circled{2} and \circled{3} share the encoding mechanism, their input and how \mechanism treats the outcome are pretty different. Step~\circled{3} neither introduces a new operation nor a procedure other than those already existing from Step~\circled{2}. Therefore, it enjoys similar benefits as Step~\circled{2}, namely high parallelization and in-memory suitability. \mechanism{} runs Step~\circled{3} every time it profiles a new read of a food sample.

%% file: Sections/3_4_similaritycheck.tex
\subsection{Step~4: Multi-Species Classification per Read} \label{subsec:similaritycheck_framework}

In this step, \mechanism takes (1) the query \hd vector (Step~\circled{3}), (2) \hdrefs (Step~\circled{2}), and (3) similarity function and its corresponding parameters (Step~\circled{1}) as inputs. To determine the specie(s) that each read belongs to, \mechanism performs a similarity check between the query \hd vector and each of the prototype \hd vectors in \hdrefs. The similarity measure can vary depending on the vector representations and encoding approaches. \mechanism allows various famous mechanisms for the similarity check, such as Hamming distance~\cite{kleyko2018classification} and dot product~\cite{ge2020classification}. Usually, this step can be implemented only with simple operations (\sect{\ref{subsec:HDC_background}}). It also enjoys high parallelization, similar to the previous steps. Although a similarity metric and its related parameters highly relate to (1) the encoding mechanism and (2) hyperparameters of the \hd space, such as representations, sparsity, and the dimension of \hd vectors, and therefore it makes sense not to let them change arbitrarily, \mechanism supports changing them in Step \circled{4} as well, without needing to re-run Steps \circled{2} or \circled{3}. This is because some studies show that different similarity metrics and thresholds may outperform others depending on your application and data for a fixed set of hyperparameters and encoding mechanisms. Therefore, if one decides to change their reference database, they may need to play with these to find the right match, and \mechanism allows such investigations. Currently, \mechanism provides a default option.

\mechanism may find out that the query \hd vector is close to one, multiple, or none of the prototype \hd vectors in \hdrefs. This variety in possible outputs differentiates \mechanism from many previous \hdc-based designs~\cite{HDC-CIM-IBM, imani2018hdna, kleyko2018classification, kleyko2016holographic, imani2020dual}. In such works, mostly due to the characteristics of applications under study, researchers always assume that (1) the query \hd vector can only belong to one of the prototype \hd vectors, and (2) the class of the query \hd vector will exist in the \am. However, none of these assumptions hold for a food profiler. One read from the food sample can be related to one, multiple, or none of the reference genomes in the original reference genome database. This is because the read sequences are mostly short strings with a reasonably high probability of existence in longer reference genome sequences. It is also not uncommon that the query \hd vector does not belong to any of the reference genomes in the initial reference genome database. This case can happen when, for example, (1) there is either an unknown species in the food sample, (2) one incorrectly excludes the corresponding reference genome in the initial reference genome database, or (3) an uncorrected sequencing error has happened. A food profiler should capture such cases. This difference between how many prototype \hd vectors in \hdrefs can be assigned to one query \hd vector is a key difference that affects both the following abundance estimation step and final results. It also distinguishes this work further from previous \hdc-based proposals for different applications. Step~\circled{4} also enjoys high parallelization and in-memory suitability features similar to previous steps.

%% file: Sections/3_5_relativeabundance.tex
\subsection{Step~5: Species Level Abundance Estimation} \label{subsec:abundanceestimation_framework}

In Step~\circled{5}, \mechanism performs a relative abundance estimation based on the results of Step~\circled{4}. This step is particularly needed for a food profiler in which one query \hd vector can be similar to one or more classes/species. \mechanism categorizes each query \hd vector into (1) uniquely-mapped, (2) multi-mapped, and (3) unmapped, taking a two-step approach. In the first step, \mechanism assigns the uniquely mapped query \hd vectors to the species they are similar to. In the second step, \mechanism assigns the multi-mapped query \hd vector to multi-species proportionally to the number of reads that have been uniquely aligned to in the first step divided by the length of species (reference genome). \mechanism's Step~\circled{5}  can be extended to support different assignment policies for the multi-mapped reads. We leave investigating the effect of such methods for future work.

%% file: Sections/4_EvaluationDemeter.tex
\section{Demeter's Evaluation} \label{sec:evaluation_Demeter}

\subsection{Methodology}  
\label{subsec:methodology_Demeter}

We implement a multi-threaded highly-parallelized version of \mechanism in \verb!C++! using SeqAn library~\cite{seqan}, called \cmechanism. SeqAn library is an open-source optimized library for biological data. \cmechanism verifies the accuracy of \mechanism. We also implement a GPU version of \mechanism, \gmechanism. \gmechanism uses CUDA streams for parallelizing data copy operation between shared memory and global memory with other computations as much as possible. It implements the similarity check using the parallel reduction technique introduced by Harris et al. \cite{harris2007optimizing} in the shared memory. All of our experiments run on a 128-core server with AMD EPYC 7742 CPUs~\cite{AMDEPYC7742CPU} and with \SI{500}{\giga\byte} of DDR4 DRAM. \gmechanism runs on an NVIDIA RTX 2080Ti GPU. Our sensitivity analysis shows that binary \hd vectors of size 40,000, with dense distributed representation (DDR~\cite{kleyko2018classification}) and \ngram-based encoding mechanism, strike a sweet spot in the tradeoff between accuracy, required memory, and performance. Therefore, unless otherwise stated, our evaluations use these setups.

\noindent 
\textbf{Accuracy Metrics.} We capture the four fundamental rates from a (food) profiler when considering the presence and absence of each species in the output, i.e., True Positive (TP), False Positive (FP), False Negative (FN), and True Negative (TN) Rate. Based on these rates, \mechanism reports two standard metrics of Precision and Recall~\cite{Micop, metalign, kraken2} to assess the accuracy of our (food) profilers. 

\noindent 
\textbf{Performance Metrics.} Performance analysis consists of three experiments: (1) Build time, (2) Query time, and (3) Query throughput or speed. This separation has two main reasons. (1) Build time is normally a one-time job and does not affect the overall profiler’s performance. Therefore, it is only fair to separate build time and query time. (2) Query time is simply the required time for profiling one single read. However, throughput is measured by million reads per minute ($\frac{MR}{m}$) and should be differentiated as it can get affected easily by other factors such as the size of the data structure, the classifier’s parallelization capability, or the infrastructure’s computation and storage/memory limitations (e.g., duplicating capabilities).

\noindent 
\textbf{Datasets.} We have two sets of datasets. (1) Genome sequences used as a reference database. (2) Genomes sequences used as food samples and input queries. We consider AFS20 and AFS31~\cite{AFSMetacache, ripp2014all} as our reference genome datasets. These datasets are two commonly used datasets consisting of 20 and 31 food-related reference genomes related to animals whose sizes vary from \SI{12}{\mega\byte} to \SI{14}{\giga\byte}. 
AFS31 is currently also the biggest reference dataset used in food profiling. Food sample reads or queries are from calibrator sausage samples from ENA project ID  PRJEB34001 \cite{PRJEB34001}, and PRJNA271645 \cite{PRJNA271645}. These reads are real short-read sequences from a mixture of food ingredients such as chicken, turkey, etc., sequenced on an Illumina HiSeq machine.

\noindent 
\textbf{Baselines.} We compare \mechanism against \metacache \cite{AFSMetacache} (the most accurate food profiler) \krakennew \cite{kraken2}, \kbr \cite{bracken}, and \clark \cite{clark}, the top 3 alignment-free and fastest metagenomic profilers that are also commonly used for food profiling.

\subsection{\mechanism's Accuracy Analysis} \label{subsec:accuracy_evaluation_Demeter}

\begin{figure}[htbp]
\centering
    \includegraphics[width=1\linewidth]{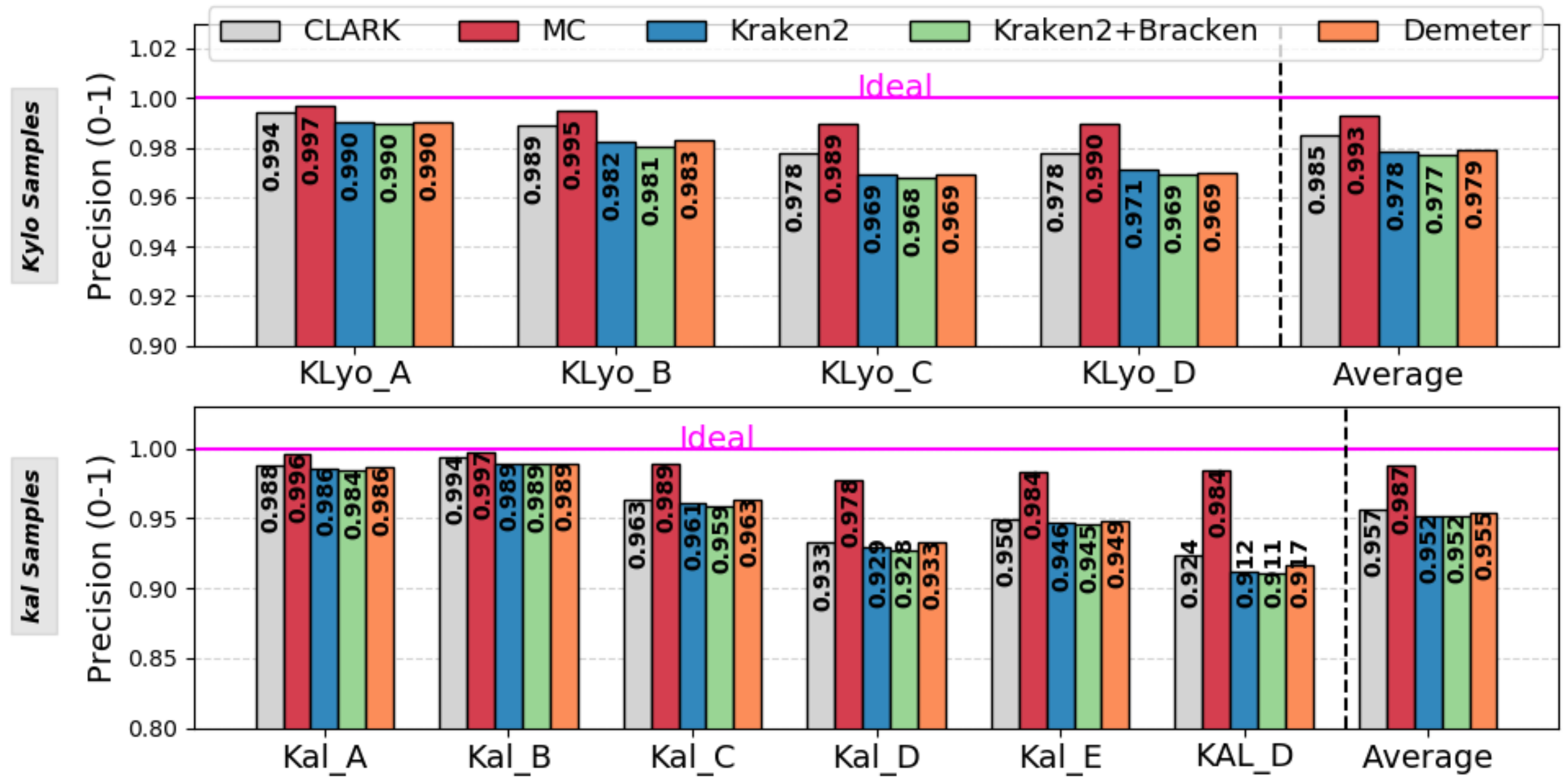}
    \caption{Precision rate for Kylo and Kal Samples on AFS20.}
    \label{fig:precision}
\end{figure}

\begin{figure}[htbp]
\centering
    \includegraphics[width=1\linewidth]{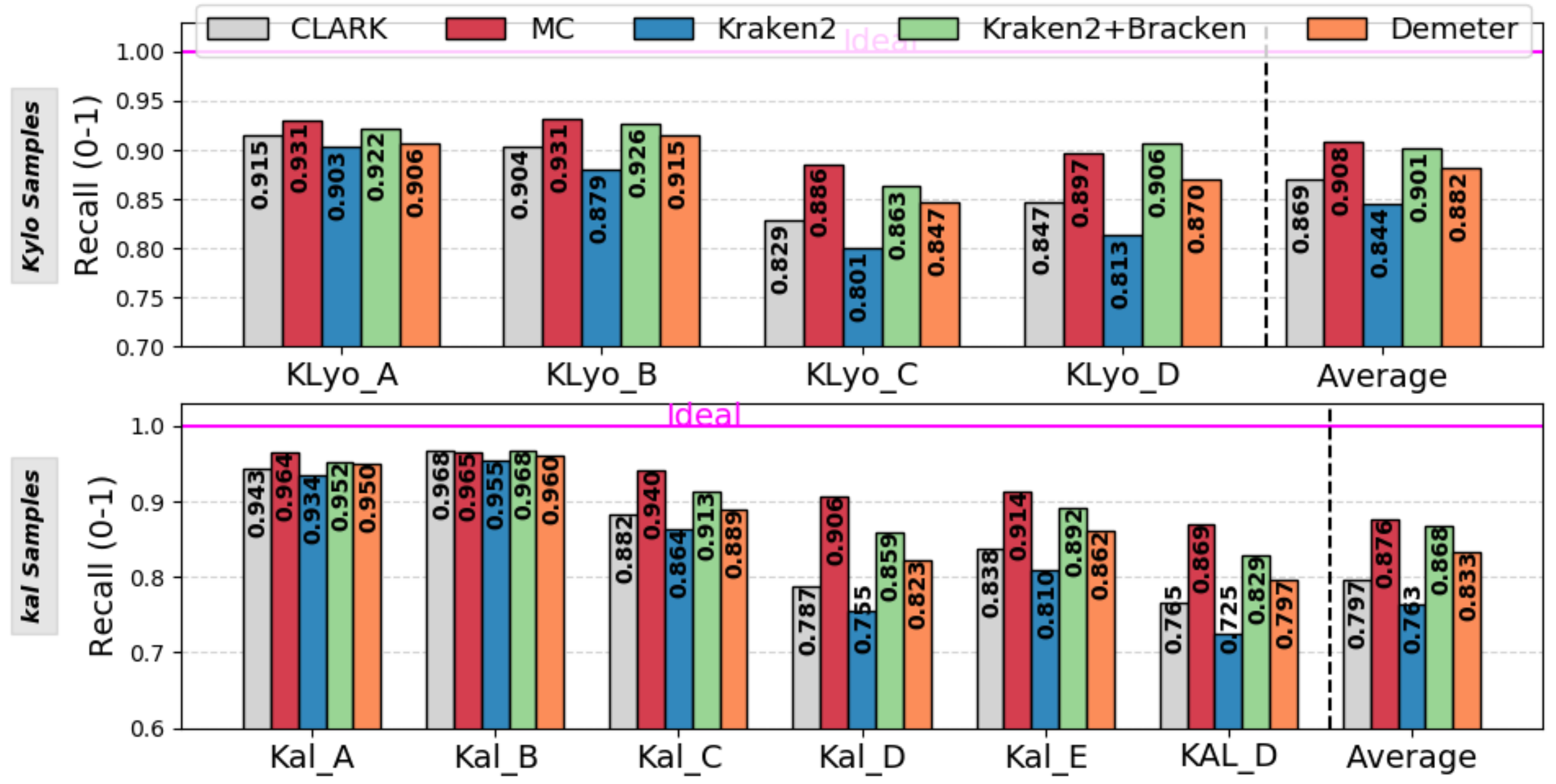}
    \caption{Recall rate for Kylo and Kal Samples on AFS20.}
    \label{fig:recall}
\end{figure}

\figs{\ref{fig:precision} and \ref{fig:recall}} present the results for the precision and recall of all evaluated food profilers on the species levels over AFS20 for Kylo and Kal food samples~\cite{PRJEB34001, PRJNA271645}, respectively. Note that the relative abundance of higher taxonomy levels is not of importance in food profiling. Additionally, those calculations highly depend on the propagation method from species level to those levels. Therefore, they have been excluded from this study. 

We observe that \mechanism stands very close to the most accurate profiler, \metacache, and has only 1.4\% and 2.6\% less precision and recall, respectively, for KLyo samples. Moreover, \mechanism achieves similar results on AFS31 and Kal samples. Note that accuracy is very much data-dependent, and indeed this accuracy drop is acceptable for a food profiler. The results of the latest comparison between current (metagenomics) profilers~\cite{meyer2021critical} show an Std error of the mean ranging from 0 to 5\% regarding the precision and recall among various widely-used profilers on different datasets. 

We conclude that \mechanism is accurate and achieves high precision and recall for food samples. These results show that \mechanism's \hdc-based classification approach followed by our abundance estimation technique does not hurt the accuracy of the profiler compared to baselines.

\subsection{\mechanism's Software Performance Analysis}  \label{subsec:performance_evaluation_Demeter}

\begin{figure}[htbp]
\centering
    \includegraphics[width=1\linewidth]{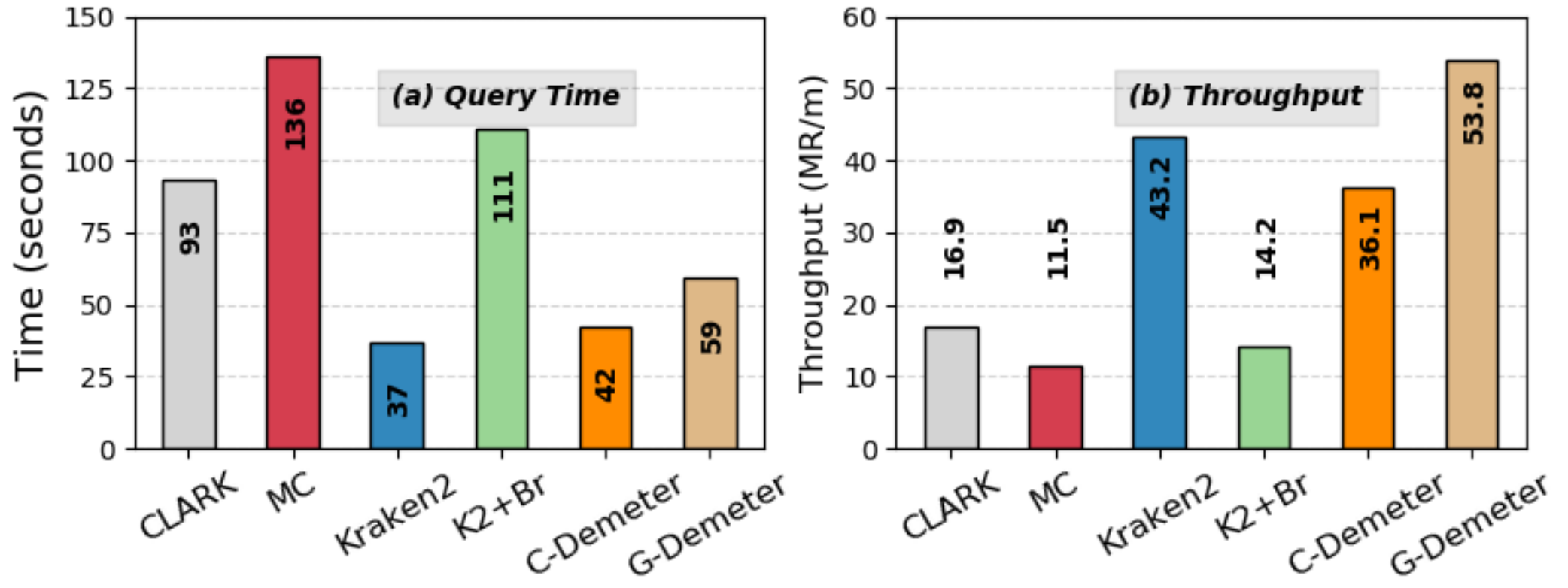}
    \caption{(a) Query time and (b) Query throughput on AFS20.}
    \label{fig:QueryAndThroughout_CPUGPU_AFS20}
\end{figure}

\begin{figure}[htbp]
\centering
    \includegraphics[width=1\linewidth]{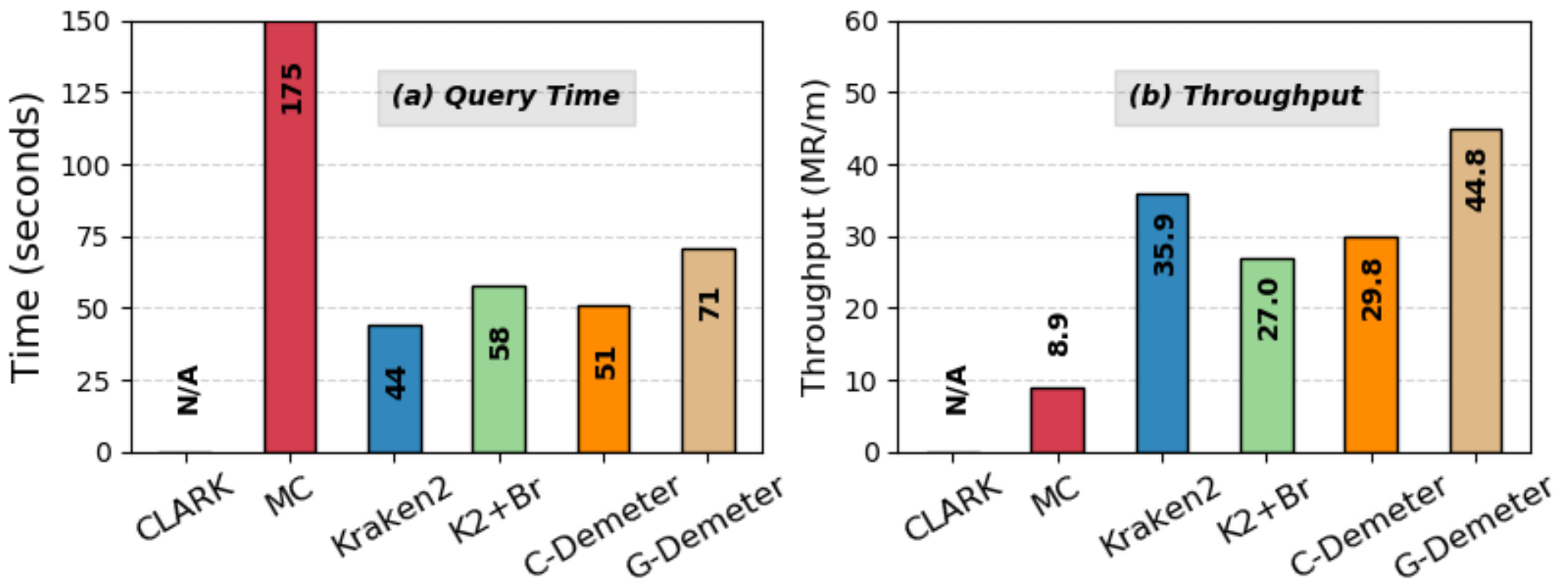}
    \caption{(a) Query time and (b) Query throughput on AFS31.}
    \label{fig:QueryAndThroughout_CPUGPU_AFS31}
\end{figure}

\fig{\ref{fig:QueryAndThroughout_CPUGPU_AFS20}}-a and \fig{\ref{fig:QueryAndThroughout_CPUGPU_AFS31}}-a present the time that each profiler takes to query one (short) read from the query food sample and classify its specie(s) over AFS20 and AFS31, respectively.

We observe that both \cmechanism and \gmechanism, whereas accurate, require higher query time compared to \krakennew. The time breakdown, using Intel VTune~\cite{IntelVTune} and cudaEvents, reveals that both implementations are memory bound, meaning there exists a significant percentage of under-utilized slots due to data access issues. 

We believe that there are two main reasons behind this problem. First, the shift operation per processed character in the encoding mechanism of \mechanism. Both of these implementations store the large \hd vectors into multiple registers. Every shift operation translates to multiple copy operations among those registers, which can become costly in terms of time and energy consumption. This is why the query time is higher than expected. Second, not all prototype \hd vectors fit in the caches. Therefore, the software versions take a few cycles to read prototype \hd vectors in batches, compare them to query \hd vector, save the results, and continue with the next batch. Note that these also put a limit on the expected throughput.

\fig{\ref{fig:QueryAndThroughout_CPUGPU_AFS20}}-b and \fig{\ref{fig:QueryAndThroughout_CPUGPU_AFS31}}-b present throughput of different profilers over AFS20 and AFS31. We make three observations. First, \cmechanism achieves a lower throughput compared to \krakennew. The reasons behind this are similar to what was discussed for its longer query time. Second, we observe that \gmechanism improves the throughput by up to 24\% (depending on the reference dataset) and therefore can be used for food profiling in the industry in the near future. Third, we observe that simply increasing the number of working threads by moving from \cmechanism to \gmechanism does not improve the throughput considerably. We ask to use the commodity GPUs to perform the food profiling to cut the cost in the short term. In the long term, we propose extending \mechanism to ASIC designs (such as those we present next) that solve the new sources of inefficiency we discussed above.

However, our analysis also shows that even a massively-parallel implementation of \mechanism, \gmechanism, does not fully utilize the parallelism offered by vector operations of \hdc classification of \mechanism, while also suffering from expensive copy-pasting among registers and its inability to perform the classification efficiently on a large vector in software.

\subsection{\mechanism's Memory Analysis}  \label{subsec:storageandMemory_evaluation}

\begin{figure}[htbp]
\centering
    \includegraphics[width=1\linewidth]{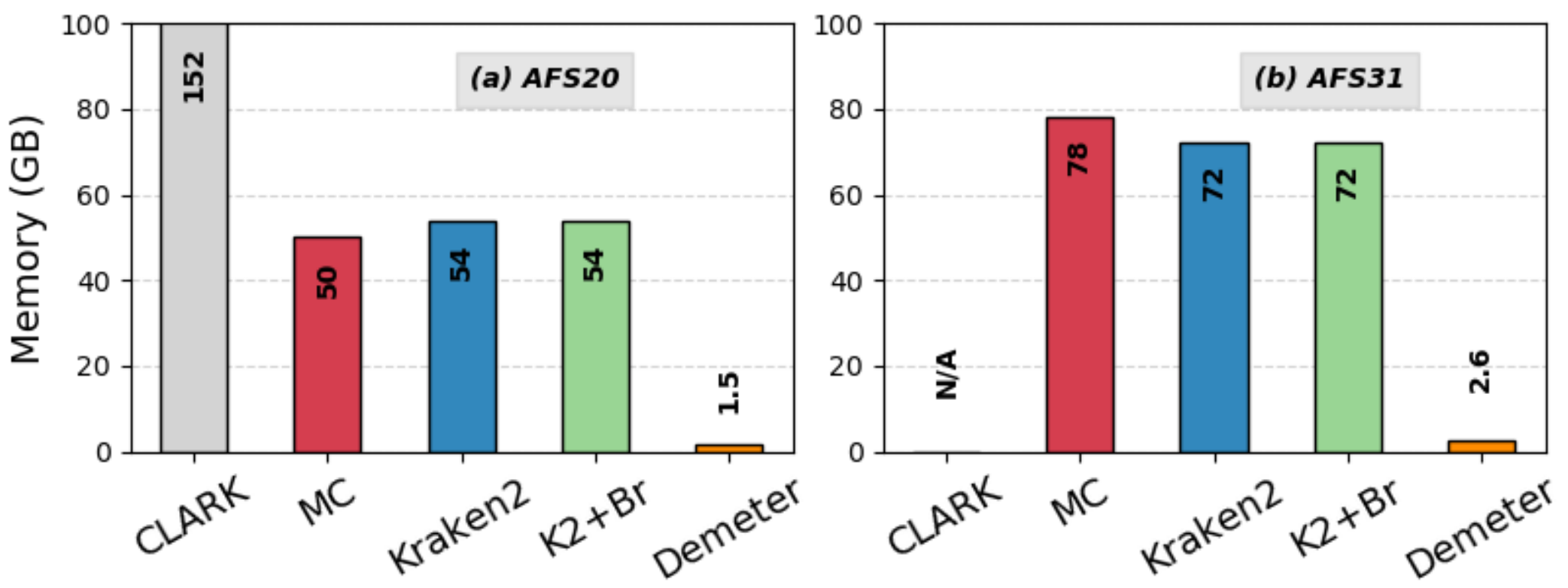}
    \caption{Required memory for (a) AFS20 and (b) AFS31.}
    \label{fig:memory}
\end{figure}

To show a key source of improvement in \mechanism (and an enabler for \accmechanism), we compared the memory requirement of \mechanism with the other food profilers. \fig{\ref{fig:memory}} presents the required memory for each profiler on AFS20 and AFS31.

We make the following two observations. First, \mechanism requires $\sim$33x and 36x less memory than \krakennew and \metacache for AFS20 database and $\sim$27x and 30x less memory for them for AFS31 database, respectively. This makes \mechanism the most efficient food profiler from a memory usage perspective. Second, the reduction in memory requirement for \mechanism is to the extent that, for the first time, the data structure of the food profiler can fit into a standard size memory and does not require a colossal RAM to manage further queries. This reduction is the primary enabler behind \accmechanism. We conclude that \mechanism is very memory efficient.

%% file: Sections/5_00_accelerator.tex
\section{\mechanism's \pim-enabled Accelerator} \label{sec:accelerator}

\mechanism is positioned as a platform-independent food profiling framework that uses \hdc. \mechanism works with large \hd vectors, is robust against errors, enjoys high parallelism, and exploits simple operations. These characteristics make \mechanism a suitable candidate for hardware acceleration. However, the interest behind accelerating \mechanism in a highly parallelizable and energy-efficient platform and specifically a \pim-enabled design goes beyond being simply its suitability and is a requisite for such a platform with two main motives.

\noindent
\textbf{Motivation 1}: As discussed in \sect{\ref{subsec:performance_evaluation_Demeter}}, a software version of \mechanism incurs a considerable cost on copy operations among registers holding intermediate \hd vectors and classification. It also performs the classification poorly due to larger than cache \hdrefs and low cache hit rate. These costs diminish all the benefits of \mechanism that come from its small data structures and memory requirement. However, one can prevent this if \mechanism is implemented in hardware as they can (1) realize the shift operation for free by only redirecting the output of each register to the next one and (2) perform the classification efficiently.

\noindent
\textbf{Motivation 2:} A software-based implementation of \mechanism still incurs a lot of unnecessary data movement for Steps~\circled{2}, \circled{3}, and \circled{4}. A hardware accelerator, especially a \pim-enabled one, can mitigate this problem greatly.

Therefore, we propose a \pim-enabled hardware accelerator for \mechanism using PCM cells. One can accelerate \mechanism using a \pim-enabled design on different memory technologies. We choose a memristor-enabled design for three main reasons. First, it is well-known that memristor-based memory technologies can perform vector-matrix multiplication~\cite{hu2018memristor, zahedi2020efficient, choi2017experimental, zahedi2021Arch} using Kirchhoff's law efficiently, making them suitable for our design. In this work, we manage to propose a hybrid row-major/column-major data mapping and intelligent data duplication scheme to perform encoding, classification, and profiling efficiently on PCM devices using this operation. Other technologies than memristors do not offer the same features for our hybrid data mapping.

Second, traditional technologies, such as non-memristor-based ones, are generally general-purpose and cost-driven. Moreover, their design does not allow even simple circuit modifications without high penalty on the area and cost. This makes them face a lot of pushback from the industry and unlikely to see future adoption. One of the advantages of memristors over them is their high density and scalability, and previous works show a wide range of accelerators using them.

Third, researchers already show the potential of accelerators based on emerging technologies for other ML-based algorithms \cite{ankit2019puma, zahedi2020efficient}. Also, multiple memory technologies already exist in current sequence machines. Therefore, it is not unreasonable to imagine one sort of these emerging memory technologies also be installed in these machines, especially for performing ML-based algorithms such as those for base-calling that are necessary for the sequencers \cite{lou2020helix}.

In this work, we focus on PCM devices, as a member of the family of memristor devices, due to our accessibility to accurate device measurements and models for these devices and leave exploring other technologies for future research.

\input{Sections/5_0_overview}

\input{Sections/5_1_IM}

\input{Sections/5_2_encoder}

\input{Sections/5_3_AM}

\input{Sections/5_4_similaritycheck_hardware}

\input{Sections/5_5_controller}

%% file: Sections/5_0_overview.tex
\subsection{Overview of \mechanism's Accelerator} \label{subsec:overview_hardware}

\begin{figure}[htbp]
\centering
    \includegraphics[width=1\linewidth]{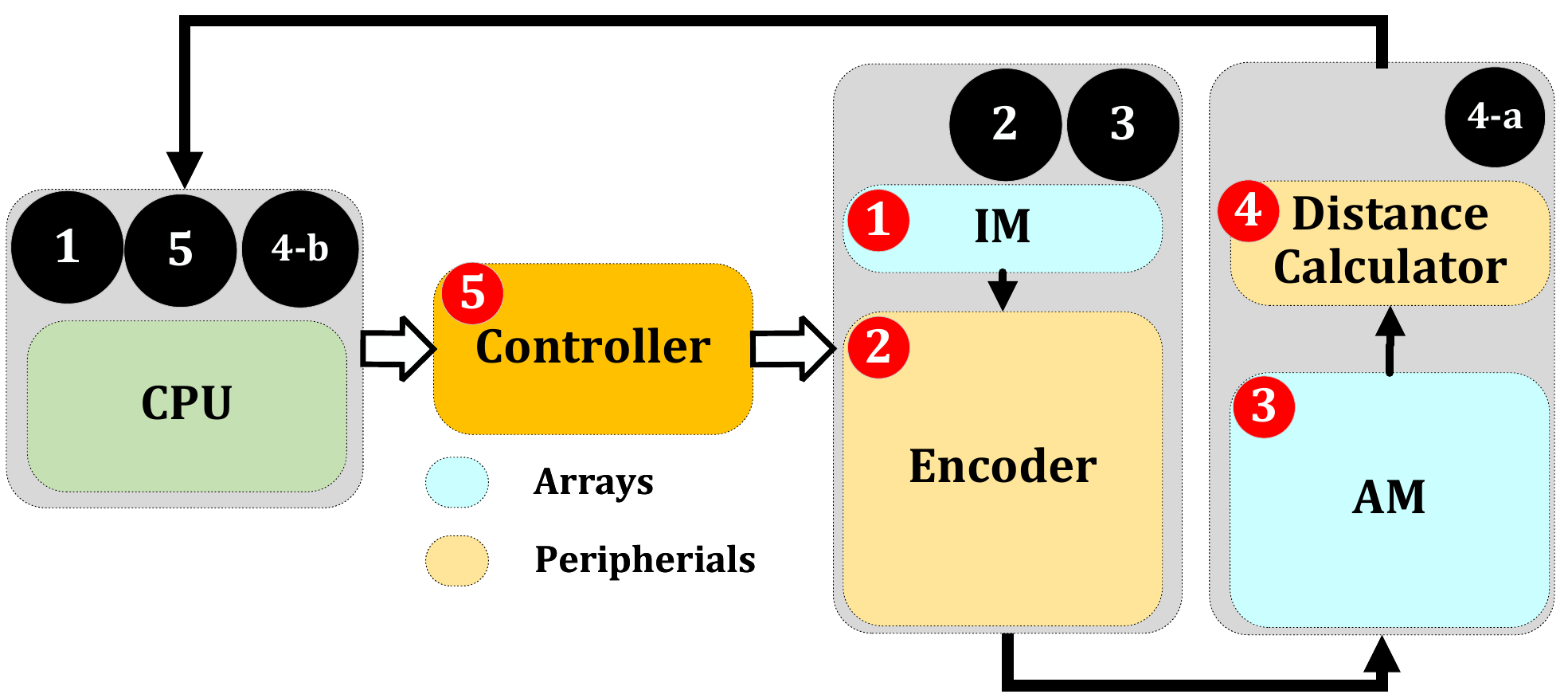}
    \caption{Overview of \mechanism's in-memory accelerator.}
    \label{fig:overview_hardware_horizontal}
\end{figure}

\fig{\ref{fig:overview_hardware_horizontal}} shows an overview of the proposed \pim-enabled hardware accelerator for \mechanism, \accmechanism. \accmechanism consists of 5 key elements: \Rcircled{1} \imlong (\im), \Rcircled{2} Encoder, \Rcircled{3} \amlong (\am), \Rcircled{4} Distance calculator, and \Rcircled{5} Controller. \im and \am units are memory units, and we implement them as PCM arrays with their control circuitry. However, the encoder and distance calculator units are computing units implemented as the periphery. The controller is a simple FSM designed to harmonize the required steps of \mechanism. The CPU initiates \mechanism by gathering the user's input (Step~\circled{1}) and then booting the controller; i.e., it sends the start command, initializes the registers, and sets the addresses to consider for food samples and/or reference genomes in the controller. In a nutshell, \accmechanism accelerates Steps~\circled{2}, \circled{3}, and \circled{4} of \mechanism. The controller returns the results of Step~\circled{4} to the CPU for final processing and performing the relative abundance estimation (Step~\circled{5}). We will discuss these units in more detail next.

%% file: Sections/5_1_IM.tex
\subsection{\imlong (\im) Design} \label{subsec:im_hardware}

We implement our \im using PCM arrays and corresponding circuits, such as decoders. \im stores the atomic \hd vectors. Binary “0” and binary “1” in an \hd vector translate to amorphous and crystalline states, respectively. In the beginning, the user (or \mechanism) generates 4 \hd vectors for each DNA alphabet in Step~\circled{1} of \mechanism and stores them in the \im. \accmechanism reads these atomic \hd vectors from \im every time it meets a new symbol. Once \mechanism fixes the \hd space, \im becomes a read-only memory. This allows us to prevent unwanted changes to the atomic vectors.

\begin{figure}[htbp]
\centering
    \includegraphics[width=1\linewidth]{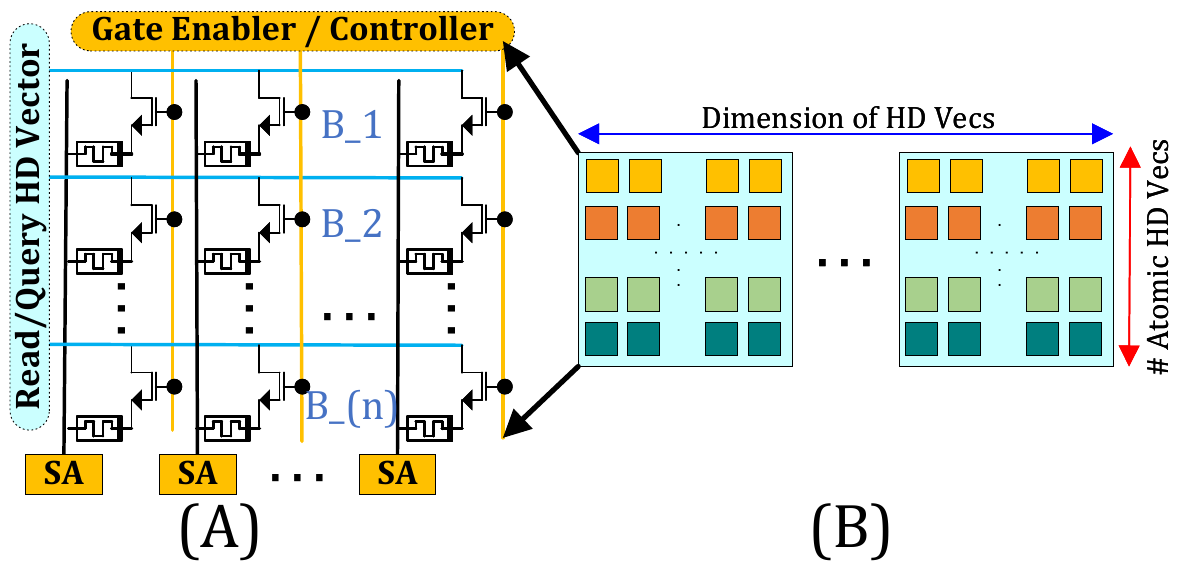}
    \caption{(A) \im design. (B) Data Mapping and placement of atomic \hd vectors in \im.}
    \label{fig:IMdesign}
\end{figure}

\fig{\ref{fig:IMdesign}}-(A) presents the \im design. The gate enabler provides access to cells that the row decoder activated. This way, the design of an entire array is achieved much easier, and the write/read disturbance effect is also mitigated to a great extent. However, this design also blocks the write on a row basis and only allows column-wise programming of \im. This does not complicate \im in any way because the atomic vectors are generated once in the beginning by the host CPU and then stored in the \im for a long time. Note that random number generators are already well-optimized in CPUs. In addition, randomly generated values inside memristors are still in early stages~\cite{diarce2014comparative, balatti2015true, puglisi2017new}, and \accmechanism can be modified later to benefit from a non-intrusive (compatible) random number generator in the future.

\fig{\ref{fig:IMdesign}}-(B) presents (1) data mapping and (2) placement of \hd vectors in the \im unit. Note that data mapping is a critical contribution of \accmechanism. \accmechanism uses a hybrid row-major and column-major data mapping for \im and \am units, respectively. \im enjoys a row-major data mapping for two reasons. First, a row-major data mapping of \hd vectors allows \accmechanism to read the cells written in one row in one cycle. This is helpful as \im is used in the encoding procedure, which is the bottleneck. Second, the used PCM model provides more \#columns than \#rows. Therefore, even if there was a method to read column cells all at once but separately, one could only store smaller chunks of an \hd vector on that column.

An important design choice regarding \im is related to the limited size of PCM arrays (512$\times$2048~\cite{HDC-CIM-IBM}). This limitation of array size (which also exists in mature memory technologies such as DRAM) prevents us from fitting an entire large \hd vector in one row or column. Therefore, one needs to break such an \hd vector into smaller chunks and store them in separate rows. Three options exist: (1) putting the chunks in the same array, (2) putting them in different arrays, (3) a hybrid approach. As shown in \sect{\ref{sec:evaluation}}, encoder is the bottleneck of our operation. Therefore, to prevent exacerbating the overhead of the encoding procedure, \im breaks a \hd vector to the largest power of two that is smaller than the number of columns available in an array (2048 in our case) and stores different chunks on different arrays. This is a direct tradeoff between the used area (\#arrays) and performance. \fig{\ref{fig:IMdesign}}-(B) also shows this placement.

%% file: Sections/5_2_encoder.tex
\subsection{Encoder Design} \label{subsec:encoding_hardware}

The encoder is the main compute unit of \accmechanism. The encoder is implemented in the periphery of arrays and executes the binding and bundling operations via a sequence of commands determined by the controller. \mechanism is capable of handling different representations (\sect{\ref{sec:framework}}). However, to reduce the complexity and make the design hardware friendly, the current design of \accmechanism only supports binary representations. In this setup, the \ngram encoding mechanism is the most common one, which \accmechanism supports. We suspect other choices are possible with the same hardware or minimal changes. We leave the exploration of those designs for future work.

\begin{figure}[htbp]
\centering
    \includegraphics[width=1\linewidth]{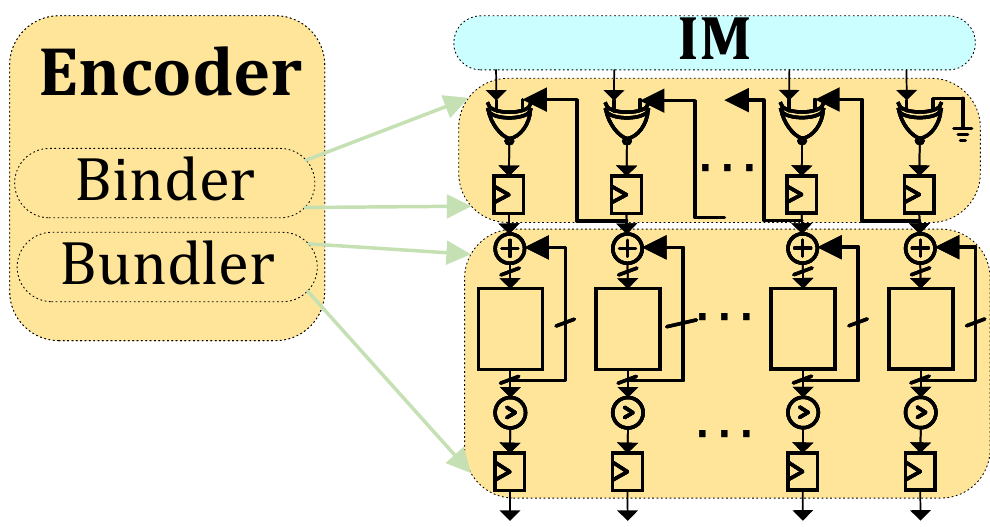}
    \caption{Encoder components and schematic.}
    \label{fig:encoder_hardware}
\end{figure}

Based on \eque{\ref{eq:ngram_xorshift}}, building an \ngram requires only simple XOR and shift operations. This bitwise XOR operation can be quickly computed after reading the atomic \hd vector from the \im with an XOR gate in the periphery. Note that one can also implement XOR using bitwise AND ($\land$) and OR ($\lor$). However, this technique requires breaking the XOR operations into minterms whose numbers increase exponentially. Any attempt to reduce them, even if empirically works as in~\cite{HDC-CIM-IBM}, will only produce approximated results and hurt the accuracy. Although some applications can tolerate such extreme accuracy loss, food profiling cannot. Note that the 2-minterm based encoding in \cite{HDC-CIM-IBM} also affects the sparsity of \ngrams (acknowledged in the paper) and limits the size of the \ngram. However, this is not the case in \accmechanism because all the operations accurately use XOR gates. This way, \accmechanism can benefit from larger \ngrams and does not hurt the density of the \hd vectors. As discussed in \sect{\ref{subsec:performance_evaluation_Demeter}}, the shift operation can quickly become a bottleneck for large \hd vectors and strings in a software-based implementation. However, this does not happen here since \accmechanism realizes the shift for free by simply redirecting each Flip-Flop (FF)'s stored value to the neighboring one every clock cycle. \fig{\ref{fig:encoder_hardware}} depicts a schematic illustration of the encoder unit.

From the hardware perspective, the encoder distinguishes the binding and bundling components completely. For the binding, the SA reads out the value from \im to one input of an XOR gate and uses the previously stored value of neighbor FFs as the second input. The encoder then stores the results in a buffer and repeats the procedure. This design choice provides \accmechanism with the cascaded logical operations, with minimum changes to the memory array, and prevents any write back and pressuring the endurance of the PCM substrates. The encoder performs this sequence N times (enforced by the signals from the controller) to build an \ngram. After it finishes creating one \ngram, it passes the \ngram to the bundler unit, resets the buffer, and starts building the next \ngram until it hits either the last character of the input or set limit per the final \hd vector.

The bundler takes \ngrams and adds them to a global \hd vector that presents each position with a counter instead of only one bit per position. It then repeats this operation for M \ngrams. Finally, the bundler applies a threshold (T) and makes a final binary \hd vector representing all the processed characters while building this vector. At this point, the encoder is done. It passes the results to be stored as prototype \hd vectors or used as query \hd vector in \am and resets both the integer-based and binary \hd vectors.

%% file: Sections/5_3_AM.tex
\subsection{\amlong (\am) Design} \label{subsec:am_hardware}

The \am unit is implemented using PCM arrays and their corresponding circuitry, similar to the \im unit. This unit takes the output of the encoding mechanism (an \hd vector) as input. Although the \am and the \im can technically be combined, \accmechanism considers separate hardware for three reasons. First, these units serve in subsequent and completely different steps in a profiling pipeline, naming encoding, and classification step. Such a distinct separation enables building a pipeline for them. Second, row-major and column-major data mapping in \im and  introduce different parallelism opportunities for encoding and classification steps of a profiling pipeline, respectively. Row-major data mapping of \im parallelizes encoding of all bits in a single \hd vector in each clock cycle. On the other hand, column-major mapping parallelizes the similarity check of one query \hd vector with all prototype \hd vectors stored vertically in that clock cycle. Third, separate hardware helps us to simplify \im design by using sense amplifiers instead of ADCs. Doing so brings various benefits in terms of area saving, energy consumption, and read-out time. Note that ADCs are usually the bottleneck of a memristor-based memory in terms of energy, area, and time~\cite{shafiee2016isaac} and that is why one only uses them when VMM or other logical operations such as Scouting Logic~\cite{xie2017scouting} are necessary.

\eque{\ref{eq:classification}} shows that for the classification, we need to count the differences between the query \hd vector and each prototype \hd vector and then decide whether or not it can belong to the corresponding class. Although one can realize this in hardware by performing XNOR operation between the two vectors followed by a pop-count operation all in the periphery, such design comes with two drawbacks: (1) it requires the pop-count operation even after the XNOR, which introduces an enormous area cost and significant delay ($log_2{D}+1$ cycles~\cite{imani2017exploring}), and (2) the \am unit, similar to the \im, only allows to write columns, not the rows. Since prototype \hd vectors are not known from the beginning (unlike atomic \hd vectors), this limitation forces us to save them all in another extra unit first and then write them back on a row basis. This is again inefficient.

However, \accmechanism proposes a new column-major data mapping and intelligent data duplication for this unit and exploits the characteristics of the PCM substrate to solve all these problems for \hdc-based classification. It is well-known that memristor-based memory technologies can perform vector-matrix multiplication~\cite{hu2018memristor, zahedi2020efficient, choi2017experimental, zahedi2021Arch}. Therefore, \accmechanism implements the required XNOR and following pop-count operations in \eque{\ref{eq:classification}} in four steps, three of which happen in the  unit and the last one in the Similarity Check unit.

\begin{figure}[htbp]
\centering
    \includegraphics[width=1\linewidth]{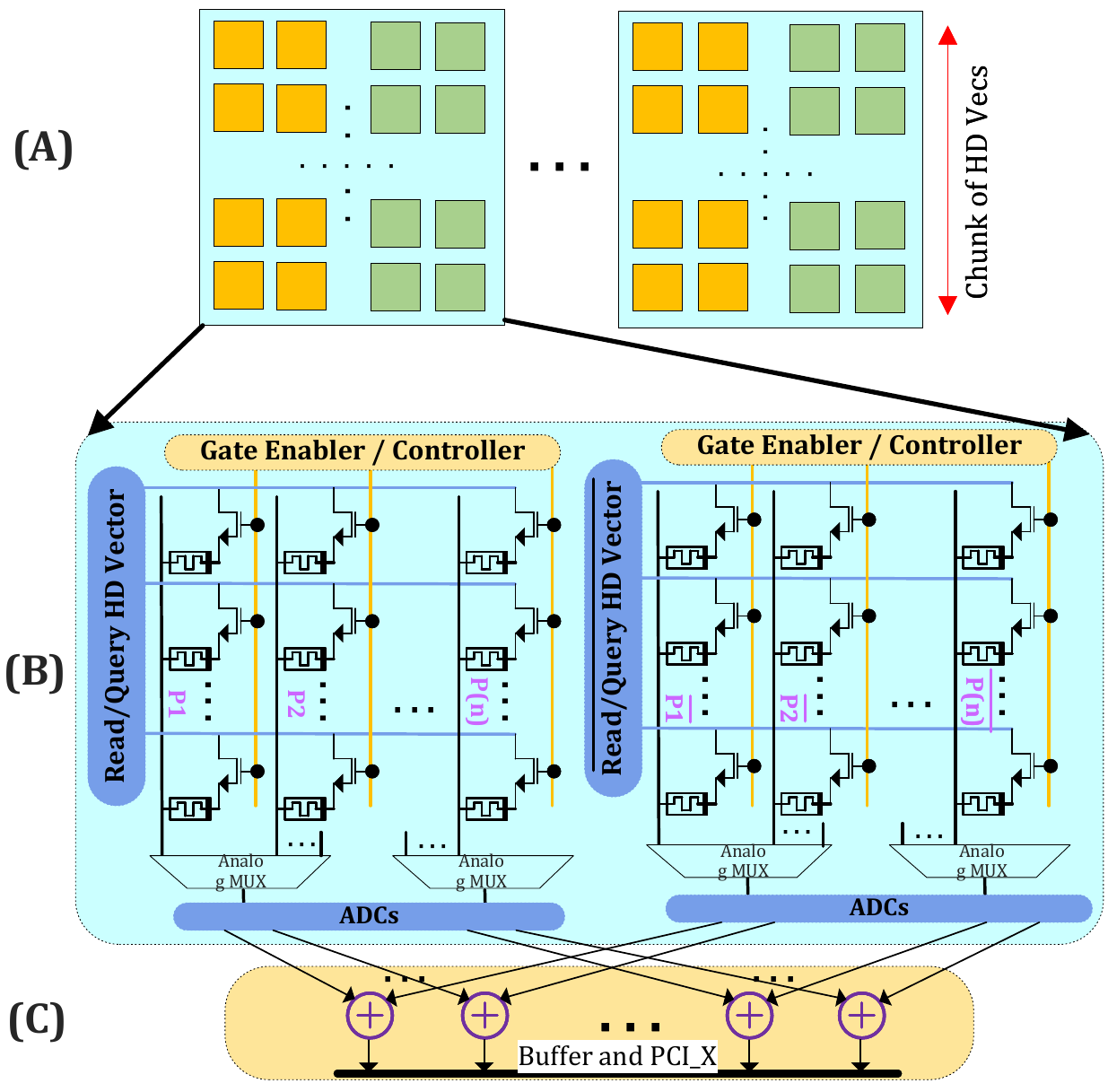}
    \caption{(A) Data mapping and placement of prototype \hd vectors in, (B) High-level  design, and (C) Partial hardware for Similarity Check unit. }
    \label{fig:AMDesign}
\end{figure}

\textbf{Step~1:} \accmechanism stores one prototype \hd vector (or a chunk of one \hd vector) in one column and its complement in the same column number of a second array. \fig{\ref{fig:AMDesign}}-A shows their placement in the \am unit. \textbf{Step~2:} \accmechanism applies the query \hd vector ($Q$) to the rows of the first array and the complement of the query \hd vector ($\bar{Q}$) to the rows of the second array with complement prototype \hd vectors (Ps), shown in \fig{\ref{fig:AMDesign}}-B. \textbf{Step~3:} \accmechanism enables columns consecutively and effectively read out the number of ones in $Q.P$ and $\bar{Q}$.$\bar{P}$ in ADCs of each array. This way, it performs two vector-matrix multiplications using Kirchhoff's law, one between Q and all Ps in the first array and one between $\bar{Q}$ and all $\bar{P}s$.  \sect{\ref{subsec:similaritycheck_hardware}} describes \textbf{Step~4} that realizes XNOR and pop-count operation simultaneously. \fig{\ref{fig:AMDesign}}-B presents a high-level illustration of \am design.

Similar to the case in \im, the limited array size of PCM substrates also prevents \accmechanism from storing a full \hd vector in one row or column of \am. To reduce the required area, and since the encoding is the bottleneck and not the classification (\sect{\ref{sec:evaluation}}), in the \am, unlike \im, \accmechanism stores the chunks of \hd vectors in the same array. \fig{\ref{fig:AMDesign}}-A takes a color-coding approach and depicts the way \accmechanism breaks prototype \hd vectors into multiple chunks and stores them in columns of \am in and among tiles. It is worth noting that \accmechanism only writes to the PCM cells once in both \im and \am units unless either the configuration file in Step~\circled{1} or the default reference genome database in Step~\circled{2} changes, as a request by the user, for example. This prevents many writes to the devices, which still have limited endurance compared to traditional memory technologies.

%% file: Sections/5_4_similaritycheck_hardware.tex
\subsection{Similarity Check Hardware} \label{subsec:similaritycheck_hardware}

The similarity check unit is a small computing unit that takes the two ADCs' output of similar columns from the two crossbars and adds them together (\textbf{Step~4}). \fig{\ref{fig:AMDesign}}-C depicts all the logic for this unit. The output of this unit is the results of XNOR and pop-count together. At this stage, the similarity check unit sends the results out to the host CPU to determine whether the similarity is close to the threshold and should be considered in the abundance estimation (\circled{4}-b, and \circled{5}). The reason behind sending the results out instead of a winner-take-all (WTA) circuit used in previous works~\cite{carvajal2000high, HDC-CIM-IBM} is two-folded. First, a WTA circuit assumes that the query matches one and only one prototype \hd vector. However, as discussed in~\sect{\ref{subsec:similaritycheck_framework} and \ref{subsec:abundanceestimation_framework}}, this is not always the case when profiling the genomics data. Second, the relative abundance estimation techniques (Step~\circled{5} in \fig{\ref{fig:overview_framework}}), although simple, require more complex and area-hungry logic circuits, which \accmechanism aims to avoid whenever possible. Therefore, since the results will be analyzed outside the PCM-substrate anyway and transferring such small data can be easily handled by interconnects between the host and \accmechanism, \accmechanism relies on the host CPU to perform the final steps of \mechanism. Note that the host is aware of prototype \hd vectors' mappings.

%% file: Sections/5_5_controller.tex
\subsection{Controller Unit} \label{subsec:controller_hardware}

The controller orchestrates all the operations of \accmechanism by generating control signals for other components. It gets the start signal and the address of food samples (or reference genomes) in the memory as its inputs. The controller outputs the results of the similarity check unit back to the host for the final steps. The controller is designed as a simple FSM machine and operates based on parameters set in Step~\circled{1}.

%% file: Sections/6_SystemIntegration_DataLayout_ISA_ProgrammingInterface_ComputationOnLargeData.tex
\section{System Integration of \accmechanism} \label{sec:system_Integration_Data_Layout_ISA_Programming_Interface}

This section discussed \accmechanism's system integration stack that enables it to operate with the host processing system.

\subsection{Address Translation}
\accmechanism works with physical addresses instead of virtual ones and is relieved of address translation challenges that exist and are dealt with in previous works~\cite{ahn2015pim, picorel2017near}. The CPU host uses the same translation lookaside buffer (TLB) lookup mechanism for normal load/store operations to translate instructions' virtual memory addresses into their physical addresses when we have a \accmechanism's instruction.

\subsection{Coherence}
\accmechanism may require modified and/or generated atomic vectors (for the \im units) or loaded prototype vectors (for the \am units). Similar to previous works~\cite{hajinazar2021simdram, holdings2010cortex, guide2011intel}, ensuring that data for \accmechanism is up-to-date is a responsibility for programmers and can be achieved easily by flushing cache lines. \accmechanism is also capable of leveraging previous \pim coherence optimizations~\cite{boroumand2016lazypim, boroumand2019conda} for further performance improvement.

\subsection{Interrupts}
We assume that the pages required by \accmechanism's \am and \im units are already present. When this is not the case, we rely on the conventional mechanisms for handling the page faults to place this data into the correct arrays. Therefore, \accmechanism does not face page fault during the execution of food profiling since pages used by \accmechanism are already loaded and pinned into \am and \im units. \accmechanism may, however, face an interrupt during a context switch. In such cases, the context of the control unit in \accmechanism will be saved and then restored when the profiler resumes.

\subsection{ISA Extensions and Programming Interface}
An expressive and efficient programming interface is a must for \accmechanism as it directly impacts the usability of \mechanism framework. To enable easy communication between \accmechanism and the programmer, we envision extending the ISA with a few instructions to let the control unit know the required operations, their timing, and where data objects reside in \im and \am units. ISA extension is possible due to the unused opcode space in the host CPU and has also been adopted in previous \pim-related architectures~\cite{seshadri2017ambit, ahn2015pim}.

\accmechanism requires 2 types of instructions: (1) \bbopinit \textit{address, size, n}: initialization of \im and \am units and (2) \bbopop \textit{size, n}: instructions for performing different operations in \accmechanism. \bbopinit is the initialization instruction informs the OS that the memory object is for \accmechanism. This way, the OS performs virtual-to-physical memory mapping required for \am and \im units. \bbopinit takes the base physical address, the size of the vector, and the intended value. For \accmechanism's operations, we extend the CPU ISA with \bbopop. \accmechanism utilizes an array-based computation model, i.e., \src and \dst are the source and destination arrays. \bbopop is the opcode, where \textit{size} and \textit{n} are \#elements in the array and \#bits in each array element, respectively. This paper assumes that the programmer will manually write suitable code for \accmechanism operations.
We summarized the required CPU ISA extensions for these operations in \tab{\ref{tab:CPUISAExtensions_ProgrammingInterface_SystemIntegration}}.

\begin{table}[htbp]
\scriptsize
\centering

\begin{tabular}{cc}
\hline
Type & ISA Format \\ \hline \hline
Initialization & \bbopinit, \textit{address}, \textit{size}, \textit{n} \\ \hline
Input Operation & \bbopop, \textit{size}, \textit{n} \\ \hline
\end{tabular}
\caption{\accmechanism ISA Extensions.}
\label{tab:CPUISAExtensions_ProgrammingInterface_SystemIntegration}
\end{table}

%% file: Sections/9_00_EvaluationAccelerator.tex
\section{\accmechanism's Evaluation} \label{sec:evaluation}

\input{Sections/9_01_MethodologyAccelerator}

\input{Sections/9_03_performanceorExecutionTime}

\input{Sections/9_05_energyandarea}

%% file: Sections/9_01_MethodologyAccelerator.tex
\subsection{Methodology} \label{subsec:methodology_accelerator}

We emulate the execution of \accmechanism using a cycle-accurate RTL model and synthesized it using UMC 65 nm technology node in Synopsys Design Compiler~\cite{synopsys}. We verify the correct behavior of our memory model using test benches and previous in-memory simulators~\cite{zahedi2020efficient, HDC-CIM-IBM}. We consider a typical operation condition of temperature 25\textdegree and voltage 1.2V when evaluating our energy consumption. All the experiments for the PCM-based \accmechanism are carried out based on PCM statistical models that capture the variations in the spatiotemporal conductivity of the devices. PCM prototypes and analytical models used for validation and further simulations are based on the results of EU project MNEMOSENE~\cite{MNEMOSENE}, led and concluded by TU Delft in 2020. \tab{\ref{tab:pcmparameters}} shows the other parameters of our PCM crossbars.

\begin{table}[htbp]
\scriptsize
\centering
\begin{tabular}{l|l}

\hline
\textbf{Technology} & PCM (512*2048 @1bit), Cell Size = 50 $F^2$\\ \hline
\textbf{Current on Conducting Devices} &  \SI{0.1}{\micro\ampere} \\ \hline
\textbf{Read Voltage} & 0.1 V \\ \hline
\textbf{Read/Write Latency} & Read=\SI{2.8}{\nano\second}, write=\SI{100}{\nano\second} \\ \hline
\textbf{ADC} & 9 bits resolution, \SI{2}{\nano\second}, \SI{4}{\pico\joule} per sample \\ \hline
\end{tabular}
\caption{PCM configuration.}
\label{tab:pcmparameters}
\end{table}

%% file: Sections/9_03_performanceorExecutionTime.tex
\subsection{\accmechanism's Performance Analysis}  \label{subsec:performance_evaluation}

This section compares the performance of SOTA profilers compared to \accmechanism, our \pim-enabled accelerator design of \mechanism.

\subsubsection{Build time.}
\begin{figure}[htbp]
\centering
    \includegraphics[width=1\linewidth]{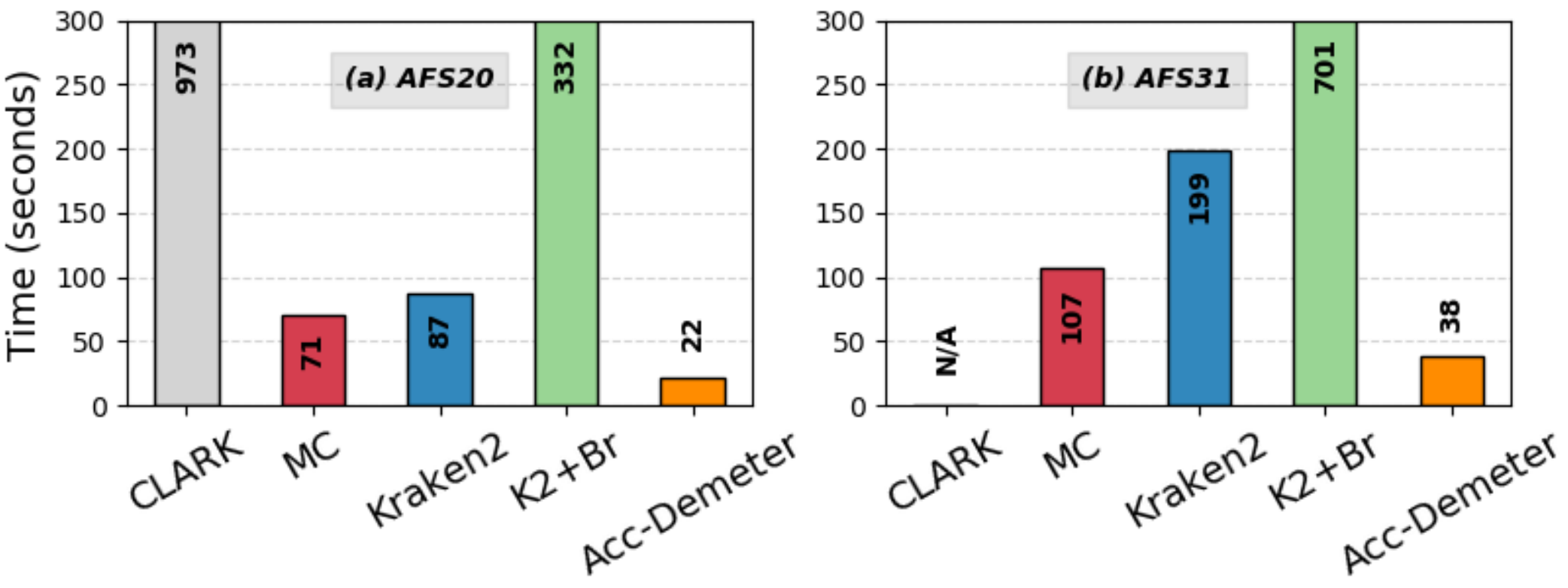}
    \caption{Build time on (a) AFS20 and (b) AFS31.}
    \label{fig:build_time}
\end{figure}

\fig{\ref{fig:build_time}} shows the build time that each profiler takes to build its initial data structure for two reference databases AFS20 and AFS31. 

We make two observations. First, \accmechanism has the lowest build time among all previous food profilers. \accmechanism builds \hdrefs corresponding to AFS20 and AFS31 $\sim$3.2x and 2.8x faster, respectively, than \metacache, the next fastest profiler. Unlike previous \hdc-based methods that are faster than their ML competitors due to the one-shot learning ability of \hdc paradigm, \accmechanism outperforms SOTA profilers due to its highly parallelized performance and simple operations being performed on \accmechanism's hardware. SOTA food profilers parse the reference genomes only once, and the one-shot learning of \mechanism is not particularly advantageous.

Second, \clark exceeds the \SI{500}{\giga\byte} memory of the system when running it for AFS31. This is in line with observations in~\cite{AFSMetacache}. Therefore, we excluded it from all analyses regarding AFS31 from now on. This case shows an excellent example of where metagenomic profilers, whereas good for lengthy and costly studies, may not be applicable for the scenario of food profiling and later food analysis and monitoring.

\subsubsection{Query time.}
\begin{figure}[htbp]
\centering
    \includegraphics[width=1\linewidth]{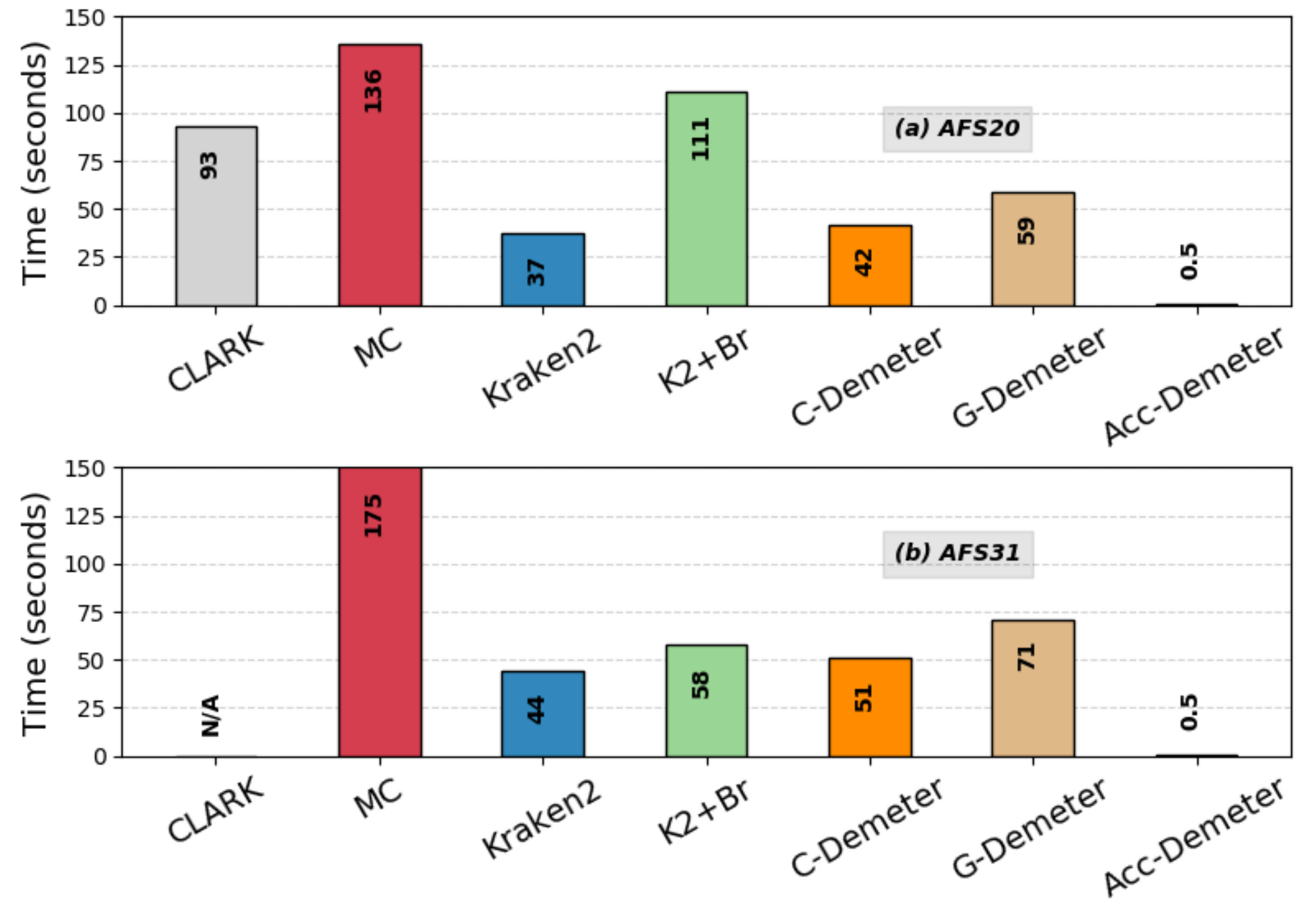}
    \caption{Query time on (a) AFS20 and (b) AFS31. }
    \label{fig:query_time}
\end{figure}

\fig{\ref{fig:query_time}} presents the time that each profiler takes to query one (short) read from the query food sample and classify its specie(s) over AFS20 and AFS31.

We make two key observations. \accmechanism improves the query time by $\sim$74x/88x and 272x/350x compared to \krakennew and \metacache, respectively, on AFS20/AFS31. This shows that the acceleration of \mechanism pays off and finally makes \mechanism not only an accurate but also a fast food profiler.

Second, the query time for \accmechanism remains almost the same for both databases and does not change much. We further investigate this and realize a bottleneck shift: Step~\circled{5} or abundance estimation that is being performed inside the CPU is now the bottleneck of \accmechanism. This happens because of the high-frequency \accmechanism achieved. However, this contrasts with other profilers that spend most of their time querying their massive data structure.

\subsubsection{Query throughput.}
\begin{figure}[htbp]
\centering
    \includegraphics[width=1\linewidth]{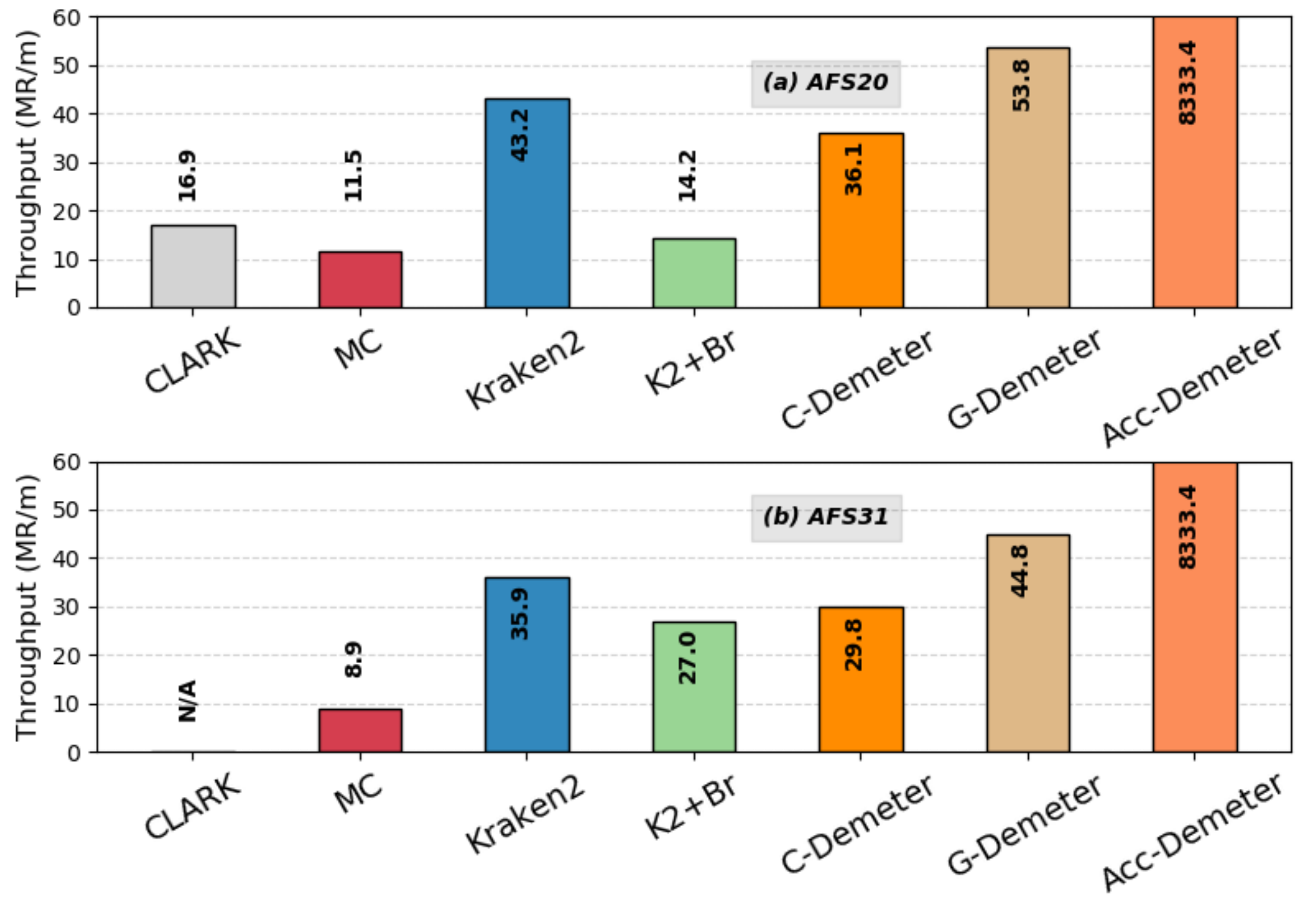}
    \caption{Query throughput on (a) AFS20 and (b) AFS31.}
    \label{fig:query_throughput}
\end{figure}

\fig{\ref{fig:query_throughput}} shows the throughput of different profilers over AFS20 and ASF31.

We make two observations. First, \accmechanism provides throughput improvement of $\sim$192x and 232x for both AFS20 and AFS31, respectively, compared to \krakennew, the second food profiler regarding throughput. This more remarkable improvement in throughput than query time results from \accmechanism's ability to classify one query read in parallel with the encoding of the following query. Note that the throughput analysis of the previous profiler does not consider the time for loading their data structure. Second, similar to the query time, throughput is almost the same regardless of the database due to the bottleneck shift. We conclude that \accmechanism significantly outperforms all four SOTA baselines for all performance metrics.

%% file: Sections/9_05_energyandarea.tex
\subsection{\accmechanism's Power and Area Analysis} \label{subsec:powerandArea_evaluation}

\tab{\ref{tab:energyareapermodule}} provides the area and energy consumption breakdown of different components in \accmechanism per query on AFS31.

 \begin{table}[htbp]
 \scriptsize
\centering
 \begin{tabular}{|c|c|c|c|c|}
 \hline
 Unit & Area ($mm^2$) & Area (\%)  & Energy ($nj$) & Energy (\%) \\ \hline \hline
 IM & 0.07 & 3.1 & 1.179E-06 & 7.4 \\ \hline
 Encoder & 1.375 & 78.3 & 1.43E-05 & 90.6 \\ \hline
 AM & 0.15 & 8.4 & 2.47E-07 & 1.56 \\ \hline
 Similarity & 0.1815 & 10.2 & 6.91E-08 & 0.4 \\ \hline
 
 \end{tabular}
 \caption{Area and power breakdown of \accmechanism.} 
 \label{tab:energyareapermodule}
 \end{table}

We make two observations. First, the logic for the encoder unit is the most energy and area hungry unit among all others, more than 90\% and 78\% energy and area of the whole \accmechanism. This is expected because (1) the encoder consists of many CMOS circuits, whereas AM and IM are small memory units with PCM technology, and (2) the encoder is in the heart of all operations in \mechanism, and we spend most of our time in this unit. We argue that this amount of logic around our array is still justifiable. Second, compared to the die area in an Intel Xeon E5-2697 CPU~\cite{fujiki2019duality}, \accmechanism only has an area overhead of less than 2\%. We conclude that \accmechanism is low-cost in terms of die area.

Our evaluations show that \accmechanism can perform a 9.45Mbp query per joule. Unfortunately, measuring the energy consumption of other profilers and having an apple-to-apple comparison between the energy consumption of this method with other ones is hard. However, Merelli et al.~\cite{merelli2018low, d2019combining} show that running \krakennew with querying an even smaller data structure built from a reduced reference genome dataset, minikraken~\cite{eyice2015sip, merelli2018low}, can incur more energy (maximum of 0.6$\frac{Mbp}{j}$). This considerable difference happens because of three reasons: (1) \krakennew queries a more complex data structure compared to \accmechanism and requires more complex operations, (2) \krakennew queries a bigger data structure for its query, and (3) \krakennew incurs significant data movement between the memory and the processing unit. All of these limitations exist in similar forms in \clark and \metacache. We conclude that \accmechanism is more energy-efficient than all four SOTA baselines.

%% file: Sections/7_LimitationOrDiscussions.tex
\section{Discussions and Future Works} 
\label{sec:discussionsandfutureworksandlimitations}

\noindent
\textbf{Capacity.} We define the capacity of \mechanism as the ratio between the number of reference genomes encoded as prototype \hd vectors to the size of \hd space for a competitive profiling accuracy target. The higher \#prototype \hd vectors are, the bigger capacity is needed, resulting in bigger \hd space and lower efficiency. Therefore, if one uses \mechanism, as is, as a metagenomics profiler, they cannot expect similar improvements compared to SOTA metagenomics profilers (e.g., \krakennew, on those datasets. We are currently investigating the additional techniques to enable \mechanism for those cases as well. However, we leave further analysis of required changes to \mechanism for supporting metagenomics profiling or other profiling studies with many reference genomes for future work.

\noindent
\textbf{Supported functions and representations.} As discussed (\sects{\ref{sec:background}, \ref{sec:framework}, and \ref{sec:accelerator}}), \accmechanism currently supports only binary representations and \ngram encoding mechanism. This design choice is made for simplicity and is based on acceptable accuracy results of the software version. We leave the hardware for other encoding mechanisms and data representations for future work.

%% file: Sections/10_related.tex
\section{Related Works} \label{sec:relatedwork}

To our knowledge, \mechanism is the first paper to propose a framework to perform food profiling using \hdc. \accmechanism is also the first hardware accelerator that enables low-cost and accurate in-memory profiling for a typical reference database in food profilers. We have already compared \mechanism and \accmechanism extensively to SOTA profilers in \sects{\ref{sec:evaluation_Demeter} and \ref{sec:evaluation}}.
This section briefly discusses previous software works for profilers (food or metagenomics), software or hardware of \hdc-based systems, and \pim-enabled accelerators.

\subsection{Metagenomic Profilers}
Several recent works propose approaches and techniques to directly or indirectly accelerate or improve the accuracy of metagenomics profiling, the first step of such studies. These works take three approaches: (1) Reducing the reference database’s size by pre-alignment filtering~\cite{FastHASH, SneakySnake} or heuristics for taxonomic classification techniques\cite{segata2012metagenomic, liu2011accurate, wood2014kraken, cuCLARK, metalign}, (2) Accelerating read alignment or assembly (only for alignment-/assembly-based profilers) on CPUs, FPGAs, or GPUs~\cite{brady2011phymmbl, daily2016parasail, banerjee2018asap, fei2018fpgasw, houtgast2015fpga, liu2013cudasw++, luo2013soap3}, (3) post-alignment/-assembly/-classification presence and abundance estimation heuristics~\cite{MetaPhlAn2, Micop, metalign}. \mechanism is categorized in the first group, taking a \hdc-based approach for the first time. However, compared to the first group, \accmechanism is much faster and has a lower cost (regarding both energy and area consumption). Note that \mechanism and \accmechanism are orthogonal to works in the third group, and their Step~\circled{5} can adapt their proposed techniques for the abundance estimation after the initial classification.

\subsection{\hdc-based Systems}
Many works exploit the \hdc paradigm for specific machine learning applications that require capturing temporal patterns. These works vary from language~\cite{imani2017low} and voice~\cite{kim2018efficient} detection to seizure detection~\cite{burrello2019hyperdimensional}. \mechanism is the first work that investigates \hdc in the realm of profiling genomics data. Although HDNA~\cite{imani2018hdna} and GenieHD~\cite{kim2020geniehd} propose to use \hdc for (partial\footnote{Neither HDNA nor GenieHD is capable of producing the exact type and location of edits between their query and the reference genome as in the typical outputs (.sam file) of a sequence aligner. Hence, the term "partial".}) sequence alignment of a single reference genome divided into multiple pieces, they never exploit it for any metagenomics or food profiling.

A few works also suggest various hardware platforms such as FPGAs, GPUs, or ASICs~\cite{HDC-CIM-IBM, imani2018hdna} to improve the performance of \hdc-based designs. \accmechanism is different from all of these designs in two important aspects. First, \accmechanism performs an exact pop-count operation in one cycle, performing two VMM in parallel and then adding the outputs of ADCs. This is in contrast to previous works~\cite{imani2018hdna, HDC-CIM-IBM, kim2020geniehd, burrello2019hyperdimensional} that perform an exact pop-count operation in $log_2{D}+1$ cycles, where D is the size of an \hd vector. Second, unlike~\cite{HDC-CIM-IBM}, \accmechanism can perform the required \hdc-based operations on long, non-sparse \ngrams (discussed in \sect{\ref{subsec:encoding_hardware}}). To our knowledge, \accmechanism is the only \pim-enabled accelerator for food profiling.

\subsection{\pim-enabled Accelerators}
Prior works also heavily investigate various forms of compute-capable memories~\cite{seshadri2017ambit, li2017drisa, ahn2016scalable}. Among these, only a few use in-memory capability for \hdc designs~\cite{3DVRRAM, HDC-CIM-IBM}. However, these works are either tuned for single tasks or capable of limited sizes for \ngrams like only up to 3-grams~\cite{3DVRRAM}, or only based on compact models from small prototypes with 256$\times$256 ReRAM arrays. \mechanism is the first work that proposes food profiling inside the memory. Theoretically, one can accelerate \mechanism using a \pim-enabled design on DRAM, SRAM, and other technologies. However, for the reasons iterated in \sect{\ref{sec:accelerator}}, \accmechanism exploits PCM to improve the performance of \mechanism.

%% file: Sections/11_conclusion.tex
\section{Conclusion} \label{sec:conclusion}

This paper introduces \mechanism, the first framework that enables profiling of food samples via \hdc whereas strictly meeting the accuracy of state-of-the-art profilers. \mechanism uses a five-step approach to enable species-level profiling using \hdc. This paper also introduces the first PCM-baed \pim-enabled hardware accelerator, called \accmechanism. We evaluate \mechanism on software and \accmechanism using a cycle-accurate model based on a small-scale PCM-based prototype. We design \mechanism and \accmechanism to (1) address the key challenge of \hdc-systems when facing a massive input, (2) eliminate the need for a powerful machine with very large memories, and (3) prevent unnecessary data movement between memory and processing units and therefore prevent wasting time and energy. We achieve significant performance and energy benefits over the SOTA CPU implementations whereas achieving the same accuracy. We hope that future work builds on top of our framework and its hardware and extends it to further improve our food profiling systems.